\documentclass[aps,prb,showpacs,twocolumn,superscriptaddress]{revtex4-1}

\usepackage{amsmath}
\usepackage{amssymb,amscd,bm,dsfont,wasysym,mathrsfs,latexsym,psfrag,accents,mathtools}
\usepackage{graphicx}
\usepackage{multirow}
\usepackage{dcolumn}
\usepackage{float}
\usepackage{color}
\usepackage{xcolor}
\usepackage{amsthm}

\newcolumntype{L}[1]{>{\raggedright\arraybackslash}p{#1}}
\newcolumntype{C}[1]{>{\centering\arraybackslash}p{#1}}
\newcolumntype{R}[1]{>{\raggedleft\arraybackslash}p{#1}}

			\newcommand{\e}[1]{\begin{align}{#1}\end{align}}	
		
		\newcommand{\f}[2]{\frac{#1}{#2}}
		\newcommand{\tf}[2]{\tfrac{#1}{#2}}
		
		\newcommand{\la}[1]{\label{#1}}

		\newcommand{\q}[1]{Eq.\ (\ref{#1})}
		\newcommand{\qq}[2]{Eqs.\ (\ref{#1}-\ref{#2})}
		\newcommand{\s}[1]{Sec.\ \ref{#1}}
		\newcommand{\fig}[1]{Fig.\ \ref{#1}}		
		\newcommand{\app}[1]{App.\ \ref{#1}}				
		\newcommand{\tab}[1]{Tab.\ \ref{#1}}
		\newcommand{\ocite}[1]{Ref.\ \onlinecite{#1}}

		\newcommand{\iwith}{\ins{with}}

		\newcommand{\eq}{=&\;}

		\newcommand{\R}{\mathbb{R}}

	\newcommand{\eikr}{e^{i\bk \cdot \br}}

\newcommand{\nabk}{\nabla_{\boldsymbol{k}}}

\newcommand{\lmt}{l^{\mt}}

\newcommand{\bt}{$BT_{\sma{\perp}}$}
\newcommand{\bts}{$BT_{\sma{\perp}}$ }

\newcommand{\var}{\varepsilon}

\newcommand\as{\;\;\;\;}

\newcommand{\hbr}{\hat{\br}}

\newcommand{\bd}{\boldsymbol{d}}

\newcommand{\bk}{\boldsymbol{k}}
\newcommand{\bkp}{\boldsymbol{k}^{{\sma{\perp}}}}

\newcommand{\bp}{\boldsymbol{p}}

\newcommand{\br}{\boldsymbol{r}}

\newcommand{\bv}{\boldsymbol{v}}

\newcommand{\bB}{\boldsymbol{B}}

\newcommand{\bK}{\boldsymbol{K}}
\newcommand{\bM}{\boldsymbol{M}}

\newcommand{\bR}{\boldsymbol{R}}

\newcommand{\bze}{\boldsymbol{0}}

\newcommand{\bPi}{\boldsymbol{\Pi}}

\newcommand{\bsigma}{\boldsymbol{\sigma}}

\newcommand{\frake}{\mathfrak{e}}
\newcommand{\frako}{\mathfrak{o}}

\newcommand{\mxx}{\mathfrak{X}^x}

\newcommand{\bmx}{\boldsymbol{\mathfrak{X}}}

\newcommand{\orb}{\boldsymbol{\mathfrak{A}}}

\newcommand{\W}{{\cal W}}
\newcommand{\A}{{\cal A}}

\newcommand{\inv}{\mathfrak{i}}
\newcommand{\mir}{\mathfrak{r}}

\newcommand\rot{\mathfrak{c}}
\newcommand\scr{\mathfrak{s}}
\newcommand\tra{\mathfrak{t}}

\newcommand{\sx}{\sigma_{\sma{1}}}
\newcommand{\sy}{\sigma_{\sma{2}}}
\newcommand{\sz}{\sigma_{\sma{3}}}

\newcommand{\tx}{\tau_{\sma{1}}}

\newcommand{\tz}{\tau_{\sma{3}}}

\newcommand{\ins}[1]{\;\;\;\;\text{#1}\;\;\;\;}

\newcommand{\ang}{\text{\AA}}

\newcommand{\cala}{{\cal A}}

\newcommand{\cald}{{\cal D}}

\newcommand{\calg}{{\cal G}}
\newcommand{\calh}{{\cal H}}

\newcommand{\calm}{{\cal M}}

\newcommand{\calo}{{\cal O}}

\newcommand{\noi}[1]{\noindent (#1)}

\newcommand{\mo}{\text{-}1}
\newcommand{\mt}{\text{-}2}
\newcommand{\minus}{\text{-}}

\newcommand{\braket}[2]{\big\langle #1 \big| #2 \big\rangle}

\newcommand{\bra}[1]{\big\langle#1\big|}
\newcommand{\ket}[1]{\big|#1\big\rangle}

\newcommand{\lin}{\notag \\}

\newcommand{\bpm}{\begin{pmatrix}}
\newcommand{\epm}{\end{pmatrix}}

\newcommand{\bal}{\begin{align}}

\newcommand{\dg}[1]{#1^{\scriptstyle{\dagger}}}

\newcommand{\sma}[1]{\scriptscriptstyle{#1}}

\newcommand{\Z}{\mathbb{Z}}

\begin{document}

\title{Revealing the topology of Fermi-surface wavefunctions from magnetic quantum oscillations}
  
\author{A. Alexandradinata}\affiliation{
Department of Physics, Yale University, New Haven, Connecticut 06520, USA}
\author{Chong Wang}\affiliation{Institute for Advanced Study, Tsinghua University, Beijing 100084, China}
\author{Wenhui Duan}\affiliation{Institute for Advanced Study, Tsinghua University, Beijing 100084, China}\affiliation{Department of Physics and State Key Laboratory of Low-Dimensional Quantum Physics, Tsinghua University, Beijing 100084, China}\affiliation{Collaborative Innovation Center of Quantum Matter, Tsinghua University, Beijing 100084, China}
\author{Leonid Glazman}\affiliation{
Department of Physics, Yale University, New Haven, Connecticut 06520, USA}

\begin{abstract}
The modern semiclassical theory of a Bloch electron in a magnetic field now encompasses the orbital magnetic moment and the geometric phase. These two notions are encoded in the Bohr-Sommerfeld quantization condition as a phase ($\lambda$) that is subleading in powers of the field; $\lambda$ is measurable in the phase offset of the de Haas-van Alphen oscillation, as well as of fixed-bias oscillations of the differential conductance in tunneling spectroscopy. In some solids and for certain field orientations, $\lambda/\pi$ are robustly integer-valued owing to the symmetry of the extremal orbit, i.e.,  they are the topological invariants of magnetotransport. Our comprehensive symmetry analysis identifies solids in any (magnetic) space group for which $\lambda$ is a topological invariant, as well as identifies the symmetry-enforced degeneracy of Landau levels. The analysis is simplified by our formulation of ten (and only ten) symmetry classes for closed, Fermi-surface orbits. Case studies are discussed for graphene, transition metal dichalchogenides, 3D Weyl and Dirac metals, and crystalline and $\Z_2$ topological insulators. In particular, we point out that a $\pi$ phase offset in the fundamental oscillation should \emph{not} be viewed as a smoking gun for a 3D Dirac metal.
\end{abstract}
\date{\today}


\maketitle

\section{Introduction}

The semiclassical Peierls-Onsager-Lifshitz theory\cite{peierls_substitution,onsager,lifshitz_kosevich} connects experimentally-accessible quantities in magnetic phenomena to Fermi-surface parameters of the solid at zero field. For example, field-induced quantum oscillations of magnetization\cite{dHvA} and resistivity\cite{SdH} have become the leading method to map out the shape of the Fermi surface of normal metals\cite{shoenberg,ashcroft_mermin} and superconductors\cite{champel_mineev} -- this phenomenology has been coined `Fermiology'.\cite{shoenberg} 

The semiclassical theory has been extended\cite{kohn_effham,rothI,blount_effham,wilkinson_semiclassical_harper} to incorporate two modern notions: a wavepacket orbiting in quasimomentum $(\bk)$ space acquires a geometric phase ($\phi_{\sma{B}}$),\cite{berry1984,mikitik_berryinmetal}  as well as a second phase ($\phi_{\sma{R}}$) originating from the orbital magnetic moment of a wavepacket around its center of mass.\cite{chang_niu_hyperorbit}  Further accounting for the well-known Zeeman coupling, $\lambda{:=}\phi_{\sma{B}}{+}\phi_{\sma{R}}{+}\phi_{\sma{Z}}$ is known\cite{rothI,rothII,fischbeck_review,Gao_Niu_zerofieldmagneticresponse} to be the complete, subleading (in powers of the field) correction to the Bohr-Sommerfeld quantization rule for nondegenerate bands.\cite{lifshitz_kosevich,onsager,zilberman_wkb} $\lambda$ is measurable as  a phase offset in oscillations of the magnetization/resistivity in 3D solids, as well as in fixed-bias oscillations of the differential conductance in tunneling spectroscopy.\cite{Sangjun_Cd3As2,Ilija_SnTe}  

While it is conventionally believed that $\phi_{\sma{B}}{=}0$ vs $\pi$ distinguishes between Schr\"odinger and Dirac systems,\cite{igor_kopelevich} we propose to view $\phi_{\sma{B}}/\pi$ as a continuous quantity that is sometimes fixed to an integer in certain space groups and for certain types of field-dependent orbits; moreover, while $\phi_{\sma{R}}$ vanishes for centrosymmetric metals without spin-orbit coupling (SOC), it plays an oft-ignored role in most other space groups. Our comprehensive symmetry analysis identifies the (magnetic) space groups in which $\lambda/\pi$ is robustly integer-valued -- we will formulate $\lambda$ as a topological invariant in magnetotransport, which is distinct from the traditional formulation of topological invariance in band insulators.\cite{schnyder2008A,kitaev2009}  We also extend our symmetry analysis to the multi-band generalization of $\lambda$, with envisioned application to bands of arbitrary degeneracy ($D$); $D{=}2$ is exemplified by spin degeneracy. 

Let us outline the organization of the main text. We begin in \s{sec:multibandquantrule} by introducing the multi-band quantization rule, and describing how $\lambda$ appears as the subleading phase correction. Experimental methods to extract $\lambda$ are discussed in  \s{sec:extractlambda}; we present here generalized Lifshitz-Kosevich formulae for the oscillatory magnetization and density of states. These formulae extend previous works\cite{lifshitz_kosevich} in their applicability to orbits of any energy degeneracy and symmetry, including orbits in magnetic solids. In \s{sec:symmetrylambda}, we provide a general group-theoretic framework to identify solids for which $\lambda$ takes only discrete values. In addition, our symmetry analysis identifies the symmetry-enforced degeneracy of Landau levels; where degeneracy is not enforced, symmetry may nevertheless constrain the possible splittings of Landau levels. We exemplify our symmetry analysis with several case studies in \s{sec:casestudies}, including: graphene, transition metal dichalchogenides, surface states of topological insulators, and 3D Weyl and Dirac metals. In particular, we point out that a $\pi$ phase offset in the fundamental oscillation should \emph{not} be viewed as a smoking gun for a 3D Dirac metal. We recapitulate our main results in the concluding \s{sec:discussion}; a final remark broadens the applicability of our symmetry analysis to matrix representations of holonomy\cite{barrysimon_holonomy} in the Brillouin torus, also known as Wilson loops\cite{wilczek1984} of the Berry gauge field.\cite{berry1984}  



\section{Multi-band Bohr-Sommerfeld quantization rule}\la{sec:multibandquantrule}

The quantization rule is derived from the effective Hamiltonian ($\calh$) that describes the low-energy dynamics of Bloch electrons in a field.\cite{kohn_effham,blount_effham,wannier_fredkin,rothI,nenciu_review} In a basis comprising $D$ field-modified Bloch functions at each wavevector, $\calh(\bK)$ is a Weyl-symmetrized, matrix function of the kinetic quasimomentum operators, whose noncommutivity is manifest in $\bK{\times}\bK{=}{-i}e\bB/c$. $\calh$ is asympotically expandable in powers of the field: $H_0{+}H_1{+}\ldots$, where the leading-order term is the  Peierls-Onsager Hamiltonian,\cite{peierls_substitution,onsager} while the subleading terms $H_1{:}{=}H_{{1}}^{\sma{B}}{+}H_1^{\sma{R}}{+}H_1^{\sma{Z}}$ respectively  encode the geometric phase, the orbital moment, and the Zeeman effect.\cite{kohn_effham,blount_effham,rothI}  

In the WKB approximation, the $D$-component vector wavefunction of $\calh$ generalizes\cite{AALG_100} the known single-component solution developed by Zilberman\cite{zilberman_wkb} and Fischbeck\cite{fischbeck_review} for a nondegenerate band.
In the absence of breakdown,\cite{cohen_falicov_breakdown,blount_effham,azbel_quasiclassical}  continuity of the vector wavefunction around a closed orbit ($\frako$)  affords us the following quantization rule:
\e{  l^2 S(E,k_z)+  \lambda_a(E,k_z) = 2\pi j+ \phi_{\sma{M}}.\la{rule3b}}
We have assumed here that the field is oriented in $\vec{z}$, such that $\frako$ is a contour of the band dispersion at 
fixed energy $E$ and $k_z$. By `closed orbit', we mean that $\frako$ does not wrap around the Brillouin torus. $\frako$ bounds a region in $\bkp{:=}(k_x,k_y)$-space with positive-definite area $S$.   $l{:}{=}(\hbar c/e|\bB|)^{\sma{1/2}}$ above is the magnetic length,  $j$ an integer, $a {\in} \Z_D{:}{=}\{0,1,\ldots,D-1\}$. The Maslov correction $(\phi_{\sma{M}})$ depends on the topology of the Fermi-surface orbit, e.g., it equals $\pi$ for orbits that are deformable to a circle,\cite{keller1958} but vanishes for a figure-of-eight orbit.\cite{AALG_breakdown} To leading order in $\lmt$, \q{rule3b} without $\lambda_a$ is a well-known result by Lifshitz and Onsager.\cite{lifshitz_kosevich,onsager,zilberman_wkb} 

The remaining term $\lambda_a$ is defined through the spectrum $(\{e^{\sma{i\lambda}_a}\}_{\sma{a=1}}^{\sma{D}})$ of the propagator ($\A$) that is generated by $H_1$ over the cyclotron period. $\cala$ may be expressed as a path-ordered exponential (denoted $\overline{\exp}$):  
\e{\A[\frako] = \overline{\exp}\left[i\textstyle{\oint}_{{\frako}} \left\{(\bmx+\orb) \cdot d\bk +  \tf{g_0\hbar}{4mv^{\sma{\perp}}}\sigma^z{|d\bk|} \right\} \right], \la{definenonabelianunitary}}
with $\frako$ carrying a clockwise orientation; the above three one-forms represent contributions by $H_{{1}}^{\sma{B}}$, $H_1^{\sma{R}}$ and $H_1^{\sma{Z}}$ respectively. The first one-form is the non-abelian Berry connection\cite{berry1984,wilczek1984} for the $D$-fold-degenerate subspace (henceforth denoted by $P$), and is defined by $\bmx(\bk)_{mn}{:=}i\braket{u_{m\bk}}{\nabk u_{n\bk}}$, with $\eikr u_{n\bk}$ the Bloch function of a band labelled by $n{\in}\Z_D$. The multi-band orbital magnetic moment is encoded in the second one-form 
\e{ \orb_{mn} \cdot d\bk =  {\sum}_{l\notin \Z_D} {\mxx_{ml}\Pi^y_{ln}} {dk_x}/{2{v}_y}+ (x \leftrightarrow y), \la{definerothoneform}}
with $m,n{\in} \Z_D$. $\bPi(\bk)_{ln}{=}i\bra{u_{l\bk}}e^{-i\bk \cdot \hbr}[\hat{H}_0,\hbr]e^{i\bk \cdot \hbr}\ket{u_{n\bk}}/\hbar$ are matrix elements of the  velocity operator,  with $\hat{H}_0$ the single-particle, translation-invariant Hamiltonian and $\hbr$ the position operator. For $n{\in}\Z_D$, $\bPi_{nn}{=}\bv$ is the velocity of each band in $P$, and  $v^{\sma{\perp}}{:}{=}(v^{\sma{2}}_x+v^{\sma{2}}_y)^{\sma{1/2}}$. $\bmx$ and $\bPi$ in \q{definerothoneform} comprise only off-block-diagonal matrix elements between $P$ and its orthogonal complement. The third one-form in \q{definenonabelianunitary} is the well-known Zeeman coupling, with $\hbar\sigma^z_{mn}(\bk)/2$ the matrix elements of the spin operator $S_z$, $g_0{\approx}2$ the free-electron g-factor and $m$ the free-electron mass. While the definition of  $\cala$ presumes a basis choice $\{u_{n\bk}\}_{\sma{n=1}}^{\sma{D}}$ within $P$, one may verify that the eigenvalues of $\cala$ are independent of this choice. By setting $D$ to $1$ in the above equations, the line integral of the three one-forms in \q{definenonabelianunitary} give $\lambda_1{:}{=}\lambda{=}\phi_{\sma{B}}{+}\phi_{\sma{R}}{+}\phi_{\sma{Z}}$ respectively, as we have introduced in the second paragraph of the paper. 

When \q{rule3b} is viewed at fixed field, the discrete energetic solutions $\{E_{a,j}\}$  correspond  to $D$ sets of sub-Landau levels; within each set labelled by $a$, the difference between two adjacent levels ($|E_{a,j+1}{-}E_{a,j}|$) is approximately $\hbar \omega_c{:=}2\pi/(l^2|\partial S/\partial E|)$ evaluated at $E_{a,j}$. When \q{rule3b} is viewed at constant  energy (e.g., the chemical potential $\mu$), the discrete solutions ($\{l^2_{a,j}\}$) correspond to values of the field where Landau levels successively become equal to $\mu$. In thermodynamic equilibrium, these are also the fields where Landau levels are suddenly depopulated with a periodicity: $l^2_{a,j+1}{-}l^2_{a,j}{=}2\pi/S(\mu)$, for each of $a{\in} \{1,\ldots,D\}$. This  results in various oscillatory phenomena from which we may extract $\lambda_a$.

\section{Generalized Lifshitz-Kosevich formulae to extract $\lambda$} \la{sec:extractlambda}

In the de Haas-van Alphen effect,\cite{dHvA} each extremal orbit ($\frako$) on the Fermi surface of a 3D metal is associated to an oscillatory contribution to the longitudinal magnetization (parallel to the field in $\vec{z}$):
\e{ &\delta{\calm} =- \f1{(2\pi)^{3/2}}\f{kT}{|\bB|} \, \f{S}{l|S_{zz}|^{1/2}}\lin
&\times \sum_{a=1}^D\sum_{r=1}^{\infty}e^{-\tf{r\pi }{\omega_c\tau}}\f{\sin\left[ r\left(l^2S{+}\lambda_a{-}\phi_{\sma{M}}\right) {\pm} \pi/4\right]}{r^{1/2} \,\text{sinh}\,(2\pi^2 rkT/\hbar \omega_c)},\la{oscmag3D}}
which is a sum of $D$ sets of harmonics.   Being valid in the degenerate ($\mu {\gg} kT$) and semiclassical ($\mu {\gg} \hbar \omega_c$) limits, \q{oscmag3D} is our generalization of the Lifshitz-Kosevich formula\cite{lifshitz_kosevich} to orbits of any energy degeneracy ($D$) and symmetry, including orbits in magnetic solids. In comparison, the commonly-employed Lifshitz-Kosevich formula with a `spin reduction factor'\cite{Dingle_I,shoenberg} is only applicable to two-fold degenerate orbits in solids with both time-reversal and spatial-inversion symmetries. All quantities on the right-hand side of \q{oscmag3D} are evaluated on $\frako$, which may be electron- or hole-like; the sign of $\pi/4$ (in the argument of the sine function) is negative (resp.\ positive) for a maximal (resp.\ minimal) orbit. $S_{zz}$ is the double derivative of $S$ with respect to $k_z$, and we have introduced Dingle's damping factor\cite{Dingle_collisions} that depends on the quasiparticle's mean free time ($\tau$). 

For $D{=}1$, the field-independent phase in the argument of the fundamental harmonic is sometimes referred to as the Onsager phase:
\e{-2\pi \gamma := \lambda-\phi_{\sma{M}} \pm \pi/4.\la{onsagerphase}}
If a Fermi surface has multiple extremal orbits, each extremal orbit additively contributes a term with the same functional form as \q{oscmag3D}. If two extremal orbits ($\frako_i$ and $\frako_{i+1}$) are symmetry-related, they contribute oscillatory terms that are identical in the parameters $\{S,S_{zz},\omega_c,\phi_{\sma{M}}\}$, but not necessarily for the $\lambda$-phase corrections. Generally, $\{\lambda^i_a\}_{a=1}^D{=}\{\pm \lambda^{i+1}_a\}_{a=1}^D$ (defined modulo $2\pi$), with the sign depending on the symmetry class of the orbit, as we will elaborate in \q{IIBconst} of \s{sec:symmetrylambda}. 

In 2D metals, the analogous oscillatory formula is
\e{\delta{\calm} {=} {-}\f1{\pi} \f{kT}{|\bB|} S \sum_{a=1}^D\sum_{r=1}^{\infty} e^{-\tf{r\pi }{\omega_c\tau}}\f{\sin[ r(l^2S{+}\lambda_a{-}\phi_{\sma{M}})]}{ \text{sinh}\,(2\pi^2 rkT/\hbar \omega_c)}\bigg|_{\mu}.\la{oscmag2D}}
The field dependence of $\mu$ is negligible in the semiclassical limit for 3D metals,\cite{shoenberg} as well as for 2D surface states of 3D solids.\cite{champel} In strictly-2D metals with a fixed particle density, field-induced oscillations in $\mu$ render the extraction of $\lambda$ implausible.



An alternative method to extract $\lambda_a$ is to measure the temperature-broadened, 3D density of states, defined as 
\e{\calg_{T}(E+\mu):= -\int_{-\infty}^{\infty}d\var\,f_{T}'(\var-\mu-E)\,g(\var).\la{defineTbroadenedDOS}}
Here, $g(E)$ is the density of states of 3D Landau levels (smoothened by the Dingle factor); $f_T'(x)$ is the derivative of the Fermi-Dirac distribution function, and approaches  ${-}\delta(x)$ in the zero-temperature limit. The oscillatory component of $\calg$ may be expressed as a harmonic expansion:
 \e{&\delta \calg(E)=   \sqrt{2\pi} \f{kT}{(\hbar\omega_c)^2} \f1{l^3|S_{zz}|^{1/2}} \lin
&\times \sum_{a=1}^{D}\sum_{r=1}^{\infty}r^{\sma{1/2}}e^{-\tf{r\pi }{\omega_c\tau}}\f{\cos\left[ r(l^2S{+}\lambda_a{-}\phi_{\sma{M}}) {\pm} \pi/4\right]}{ \text{sinh}\,(2\pi^2 r kT/\hbar \omega_c)}\bigg|_{E,\bar{k}_z}.\la{oscillatorydIdV}}
This quantity may be accessed via the scanning tunneling microscope (STM)\cite{Sangjun_Cd3As2,Ilija_SnTe,satya_na3bi} or by planar tunneling junctions.\cite{ivar_planartunneljunction,sun_planartunnel_SmB6} The oscillatory component of the differential conductance ($dI/dV$), averaged over the surface of a 3D metal at fixed bias voltage ($V$) for the STM, is directly proportional to 
$\delta \calg(\mu+eV)$ in the absence of surface-localized states;\cite{stm_chen}  for the planar tunneling junction, no spatial averaging is needed for a sufficiently large junction size. The just-mentioned proportionality presupposes that: (a) the tunneling matrix elements and the density of states of the STM tip are featureless in the energy range that is accessed by the bias voltage, and that (b) the tip and sample have thermally equilibrated.\cite{stm_chen,ivar_planartunneljunction}  Landau-level spectroscopy via the scanning tunneling microscope has already been reported,\cite{Sangjun_Cd3As2,Ilija_SnTe,satya_na3bi} but we are not aware that the phase offset of the oscillations has ever been measured. Further details on the derivation of \qq{oscmag3D}{oscillatorydIdV} are provided in \app{app:dhva}. 







Let us discuss how to extract $\lambda$ from dHvA data. For simplicity in presentation, we consider the magnetization oscillations contributed by  a single  orbit (extremal orbit in 3D metals). We assume that $l^2S$, $\hbar \omega_c$ and the Dingle lifetime $\tau$ have already been extracted by standard techniques.\cite{dhva_Dhillon} Let us first consider  either $\omega_c\tau {\ll} 1$ or $kT {\gg} \hbar \omega_c$, such that the harmonic expansion is dominated by the fundamental ($r{=}1$) harmonic. If $D{=}1$, then the experimental data may directly be fitted to a single sine function offset by the Onsager phase [cf.\ \q{onsagerphase}]. If $D{=}2$ (e.g., spin degeneracy), the sum of two fundamental harmonics produces an equi-frequency harmonic proportional to\cite{rothII} 
\e{ 2\bigg|\cos\bigg(\f{\lambda_1{-}\lambda_2}{2}\bigg)\bigg| \sin\left(l^2S{+}\Theta {-}\phi_{\sma{M}} {\pm} \f{\pi}{4}\right)\la{oscmag3DD2}}
with 
\e{ \Theta:= \f{\lambda_1{+}\lambda_2}{2}+\pi \,\bigg(1-\text{sign}\bigg[\cos\bigg(\f{\lambda_1{-}\lambda_2}{2}\bigg) \bigg]\bigg) \la{defineTheta}}
defined to be invariant (modulo $2\pi$) under $\lambda_j {\rightarrow} \lambda_j{+}2\pi$; these formulae will be applied to a case study of Bi$_2$Se$_3$ in \s{sec:bi2se3}. If $|S_{zz}|$ is not otherwise measurable (for 3D metals), it is not possible to fully determine $\lambda_{1,2}$ owing to our ignorance of the amplitude  of the fundamental harmonic; measuring the phase offset of the fundamental harmonic merely determines  $\Theta$, a suitably-defined \emph{average} of $\lambda_{1,2}$. $\Theta$ alone does not completely characterize the non-abelian transport within the two-band subspace. 


In the interest of measuring individual values of $\lambda_{1,2}$, we propose higher-field dHvA measurements of cleaner samples ($\omega_c\tau$ not ${\ll} 1$) and at lower temperatures ($kT$ not ${\gg} \hbar \omega_c$). In this regime, not just the fundamental but also higher ($r{>}1$) harmonics are needed to accurately represent the dHvA data.\cite{dhva_Dhillon} Since $\{\lambda_a\}_{a{=}1}^D$ is encoded in the interference of multiple harmonics, $\{\lambda_a\}_{a{=}1}^D$ may be extracted without knowledge of the absolute amplitude of a single harmonic.

In metals with multiple extremal orbits, there may exist field orientations where all orbits are related by symmetry, which simplifies the fitting to the Lifshitz-Kosevich formula.\cite{gfactorBismuth_cohen_blount} Independent of the field orientation, there exists one simplification for any non-magnetic metal: due to time-reversal ($T$) symmetry, the set of all $\{\lambda\}$ comprises only pairs that are invariant under inversion about zero ($\lambda{\rightarrow}{-}\lambda$), which effectively halves the independent parameters that require fitting. We quote this result here to exemplify the utility of a symmetry analysis of $\{\lambda\}$, which we explore in greater generality in the next section [\s{sec:symmetrylambda}]; the above mentioned constraint by $T$ symmetry is elaborated subsequently in \s{sec:globalsumrule}.

\section{Symmetry analysis of the $\lambda$ phase}\la{sec:symmetrylambda}

In certain (magnetic) space groups,  $\lambda$ [or ${\sum}_{\sma{a=1}}^{\sma{D}}\lambda_a/\pi$ for $D{>}1$] is integer-valued owing to the symmetry of the extremal orbit. To identify these orbits and space groups, it is useful to distinguish ten symmetry classes for closed orbits -- to each class we associate certain  constraints for the propagator $\cala$ and its spectrum [$\{\lambda_a\}_{a=1}^D$], as summarized in \tab{tab:tenfold}. The goal of this section is  identify the relevant symmetry class of closed orbit  -- for any physical system one chooses to study. Once this identification is made, the resultant symmetry constraints on $\{\lambda_a\}_{a=1}^D$ may be read off from the last column of \tab{tab:tenfold}, and verified experimentally from any of the methods detailed in \s{sec:extractlambda}. This will be exemplified for several case studies in \s{sec:casestudies}. 

 Let us first restate the problem in simple terms: the dynamics of Bloch electrons immersed in $\vec{z}$ are restricted  to Brillouin two-tori (\bt) of fixed $k_z$. For a $D$-fold degenerate band subspace with dispersion $\var(\bk)$, semiclassical motion occurs along (assumed) closed orbits defined by $\var(\bkp,k_z){=}E$, with $\bkp$ parametrizing \bt$(k_z)$. If multiple disconnected orbits exist within the same \bt, we assume they are sufficiently separated in $\bkp$-space that tunneling is negligible. Neglecting the subleading term ($H_1$) in the effective Hamiltonian, all Landau levels are at least $D$-fold degenerate owing to the Onsager-Lifshitz quantization rule; here and henceforth, the `degeneracy of a Landau level' is defined in units where the extensive degeneracy of a Landau level (associated to a single spinless orbit) equals $1$ [cf.\ \q{LLdeg}]. For a subset of Landau levels, this zeroth-order degeneracy is enhanced to $LD$ if a symmetry ($g$) constrains $L$ disconnected orbits to have identical shape. We may ask if (and how) $H_1$ splits this $LD$-fold (or $D$-fold) degeneracy;\footnote{Where $L{>}1$, we consider $L$ symmetry-related propagators $\A_i$ with an additional index $i{\in} \Z_L$; each of $\A_i$ is a matrix of dimension $D$, and its eigen-phases are denoted by $\lambda_a^i$, with $a\in \Z_D$.} if $L{=}D{=}1$, we ask if $H_1$ shifts the zeroth-order Landau-level spectrum at all. The answer to these questions depends on the class of symmetric orbit, which we proceed to analyze in full generality. 


\subsection{Tenfold classification of symmetric orbits}\la{sec:tenfoldorbits}

Since lattice translations trivially constrain $\cala$, we shall henceforth focus on symmetries ($g$) of the solid that correspond to nontrivial elements in the (magnetic) point group ($P$) of the solid; examples include (screw) rotations, (glide) reflections, spatial inversion and time reversal. Technically speaking, magnetic point groups differ from point groups in that the former includes a symmetry that reverses time; however this distinction is irrelevant to the following classification of symmetric orbits, hence we hereafter use `point group' democratically.

We are interested only in $g$  that maps \bts to itself; such $g$ correspond to a subgroup ($P_{\perp}$) of $P$ that generally depends on the field orientation as well as $k_z$. Any configuration of closed orbits in \bts may be divided into a disjoint set of elementary orbits $\{(g,\calo_i)\}$, where  $\calo_i$ is defined to be the smallest, closed orbit configuration that is invariant under $g$. By `invariance', we mean that for every $\bkp {\in} \calo_i$, the map of $\bkp$ under $g$ (denoted as $g {\circ} \bkp$) belongs also in $\calo_i$. Similarly, if a closed orbit $\frako {\in} \calo_i$, so would $g{\circ}\frako{\in}\calo_i$.  

There are three topologically distinct mappings of $\bkp{\in}\frako$. The simplest is the identity map, where each $\bkp$ in \bt (but not necessarily in the entire 3D torus) is individually invariant under $g$. Such mappings are labelled as class I, and all other mappings are of class II. We further distinguish between class-II mappings where $g\circ \frako$ is identical to $\frako$ up to orientation [class II-A], or they are disconnected orbits [class II-B].



There are two classes of class-I elementary orbits distinguished by whether $g$ is purely a spatial transformation, or otherwise includes a time reversal. We introduce a $\Z_2$ index $s(g)$ which equals $0$ in the former, and $1$  in the latter. Class $[I,s{=}0]$ is exemplified by \bt being a mirror/glide-invariant plane, and $[I,s{=}1]$ by $g{=}T\inv$, which is the composition of time reversal ($T$) with spatial inversion ($\inv$); all class-I symmetries are order two. Class-II elementary orbits are likewise distinguished by whether $g$ inverts time; they are additionally distinguished by whether $g$ acts on $\bkp$ as a two-dimensional rotation ($u{=}0$), or as a two-dimensional reflection ($u{=}1$). Equivalently, given that $\frako$ is clockwise-oriented, $u(g)$ distinguishes between symmetries that preserve ($u{=}0$) or invert ($u{=}1$) this orientation. In each of II-A and II-B, there are then four classes of elementary orbits distinguished by $s,u{\in}\Z_2$. This gives ten classes of elementary orbits in total, whose defining characteristics are summarized in the first three columns of \tab{tab:tenfold}.


\begin{table}[ht]	
\centering
\scalebox{0.85}{		
\begin{tabular} {|r|c|c|l|l|c|} \cline{2-6}
			
\multicolumn{1}{c}{} &\multicolumn{1}{|c}{$u$}&  \multicolumn{1}{|c}{$s$} &  \multicolumn{2}{|c}{Symmetry constraints}  & \multicolumn{1}{|c|}{$\lambda$}  \\  \hline \hline 
		  
(I)$\;\;\, \forall \, \bkp,$ & $0$& $0$ &  $\cala{=}\bar{g}\cala\bar{g}^{\mo}$    &  $\bar{g}^{\sma{2}}{=}e^{\sma{i\pi F\mu \minus i\bk \cdot \bR}}$ & $-$      \\ \cline{2-6}			 

$\bkp{=}g{\sma{\circ}}\bkp$ & $0$& $1$ &  $\cala{=}\bar{g}\cala^*\bar{g}^{\mo}$    & $(\bar{g}K)^{\sma{2}}{=}e^{\sma{i\pi F\mu \minus i\bk \cdot \bR}}$  & $e^{\sma{ i{\sum}_a\lambda_a}}{\in} \R$     \\ \hline

\multicolumn{1}{|l|}{(II-A)} & $0$& $0$ &  $\cala{=}\bar{g}\cala\bar{g}^{\mo}$      &     $\bar{g}^{\sma{N}}{=}\cala^{\sma{{\pm} N{/}L}}e^{\sma{i\pi F\mu}}$ & \multicolumn{1}{c|}{$-$}       \\ \cline{2-6}			 
			 
$\bkp \in \frako,$  & $0$& $1$ &  $\cala{=}\bar{g}\cala^*\bar{g}^{\mo}$    &    $(\bar{g}K)^{\sma{N}}{=}\cala^{\sma{\pm N/L}}e^{\sma{i\pi F\mu}}$ &  $e^{\sma{ i{\sum}_a\lambda_a}}{\in} \R$     \\ 			 \cline{2-6}

$|\frako| {=} |g{\sma{\circ}}\frako|$ & $1$& $0$ &  $\cala {=}\bar{g}\cala^{\mo}\bar{g}^{\mo}$   &  $\bar{g}^{\sma{N}}{=}e^{\sma{i\pi F\mu \minus i\bk \cdot \bR}}$  & $e^{\sma{ i{\sum}_a\lambda_a}}{\in} \R$    \\ \cline{2-6}			 

 & $1$& $1$ &  $\cala {=}\bar{g}\cala^{t}\bar{g}^{\mo}$     &  $(\bar{g}K)^{\sma{N}}{=}e^{\sma{i\pi F\mu \minus i\bk \cdot \bR}}$& \multicolumn{1}{c|}{$-$}       \\ \hline

\multicolumn{1}{|l|}{(II-B)} & $0$& $0$ &  $\cala_{\sma{i+1}}{=}\bar{g}_{\sma{i}}\cala_{\sma{i}}\bar{g}_{\sma{i}}^{\mo}$   &  $\bar{g}_{\sma{N}}\ldots \bar{g}_{\sma{1}}{=}e^{\sma{i\pi F\mu \minus i\bk \cdot \bR}}$ & $\{\lambda^{\sma{i{+}1}}_{\sma{a}}\}{=}\{\lambda^{\sma{i}}_{\sma{a}}\}$     \\ \cline{2-6}			 
			 
$\bkp \in \frako,$ & $0$& $1$ &  $\cala_{\sma{i+1}}{=}\bar{g}_{\sma{i}}\cala_{\sma{i}}^*\bar{g}_{\sma{i}}^{\mo}$    &  $\bar{g}_{\sma{N}}K\ldots \bar{g}_{\sma{1}}K{=}e^{\sma{i\pi F\mu \minus i\bk \cdot \bR}}$ & $\{\lambda^{\sma{i{+}1}}_{\sma{a}}\}{=}\{\minus\lambda^{\sma{i}}_{\sma{a}}\}$    \\ \cline{2-6}			 

$|\frako| {\neq}  |g{\sma{\circ}}\frako|$ & $1$& $0$ & $\cala_{\sma{i+1}}{=}\bar{g}_{\sma{i}}\cala^{\mo}_{\sma{i}}\bar{g}_{\sma{i}}^{\mo}$     &     $\bar{g}_{\sma{N}}\ldots \bar{g}_{\sma{1}}{=}e^{\sma{i\pi F\mu \minus i\bk \cdot \bR}}$  & $\{\lambda^{\sma{i{+}1}}_{\sma{a}}\}{=}\{\minus\lambda^{\sma{i}}_{\sma{a}}\}$   \\ \cline{2-6}			 

										& $1$& $1$ &  $\cala_{\sma{i+1}}{=}\bar{g}_{\sma{i}}\cala^t_{\sma{i}}\bar{g}_{\sma{i}}^{\mo}$    &   $\bar{g}_{\sma{N}}K\ldots \bar{g}_{\sma{1}}K{=}e^{\sma{i\pi F\mu \minus i\bk \cdot \bR}}$   & $\{\lambda^{\sma{i{+}1}}_{\sma{a}}\}{=}\{\lambda^{\sma{i}}_{\sma{a}}\}$  \\ \hline

\end{tabular}
}		
\caption{The first three columns distinguish between ten classes of elementary orbits. The map of $\bkp$ under $g$ is $g{\circ}\bkp{=}(-1)^{s(g)}\check{g}^{\sma{\perp}}\bkp$, with $\check{g}^{\sma{\perp}}$ a two-by-two orthogonal matrix  that represents the point-group component of $g$ in the plane orthogonal to the field; $s(g){=}0$ if $g$ is purely a spatial transformation, and ${=}1$ if $g$ inverts time. $|\frako| {=} |g{\sma{\circ}}\frako|$ indicates that $\frako$ is mapped to itself under $g$, modulo a change in orientation. $u(g)$ distinguishes between proper and improper transformations on $\bkp$: $(-1)^{\sma{u}}{:=}$det$\,\check{g}^{\sma{\perp}}$. Fourth and fifth columns describe how  unitary matrices $\bar{g}$ (that represent the symmetry $g$) constrain the propagator. Column six summarizes the constraints on $\lambda_a$; if there are none, we indicate this by $-$.   
	\label{tab:tenfold}}
\end{table}

In class I and II-A, $\calo_i$ is composed of a single orbit $\frako$ which is self-constrained by $g$. In II-B, $\calo_i$ is composed of $L$ disconnected  orbits which are mutually constrained as $g{\circ}\frako_i{=}(-1)^{\sma{u}}\frako_{i+1}$ and $\frako_{i+L}{:=}\frako_i$. To clarify, $\frako$ and $\{\frako_j\}_{\sma{j{=}1}}^{\sma{L}}$ are all clockwise-oriented, and $-\frako_i$ denotes an anticlockwise-oriented orbit.  $L$ was introduced in the second paragraph of  \s{sec:symmetrylambda}, and is more precisely defined here as the smallest integer for which $g^L{\circ}\bkp{=}\bkp$ for all $\bkp$; generally, $L$ divides the order ($N$) of $g$, e.g., $L{=}3$ and $N{=}6$ for the composition of $T$ and a six-fold rotation.

\subsection{Symmetry constraints on the propagator $\cala$}

Column four summarizes how $g$ constrains $\cala$ [I,II-A] and $\{\cala_j\}$ [II-B], which are respectively the propagators for the self-constrained $\frako$ and mutually-constrained $\{\frako_j\}$. The corresponding spectra of the propagators are denoted as $\{e^{i\lambda_a}\}_{\sma{a=1}}^{\sma{D}}$ and $\{e^{i\lambda^j_a}\}_{\sma{a=1}}^{\sma{D}}$. 

The unitary matrices $\bar{g}$ that constrain these propagators form a projective representation\cite{Cohomological,shiozaki_review} of the point group $P_{\perp}$, as summarized in column five. Any $g$ that inverts time $[s{=}1]$ has the antiunitary representation $\bar{g}K$, with $K$ implementing complex conjugation; otherwise $[s{=}0]$, $g$ has the unitary representation $\bar{g}$. The relations in column five are closely analogous to the space-group relations\cite{Lax} satisfied by $g$:
\e{ g^N{=}\frake^{\mu}\tra_{\sma{\bR}}, \iwith \mu(g){\in}\{0,1\}. \la{spacegrouprel}}
Here, $N(g){\in}\mathbb{N}$ is the smallest integer such that $g^N$ is a translation ($\tra$) by the lattice vector $\bR$, possibly composed with a $2\pi$ rotation (denoted $\frake$). Note the similarity in definition of $N$ with $L$, as was defined in \s{sec:tenfoldorbits}; in general, $L$ divides $N$. $\bR$ is nonzero for nonsymmorphic symmetries such as screw rotations and glide reflections;\cite{Lax} the translation $\tra_{\sma{\bR}}$ is represented on Bloch functions with wavevector $\bk$ by the phase factor $e^{{-}i\bk \cdot \bR}$ [cf.\ column five].  $\frake$ in \q{spacegrouprel}  is represented in column five by a phase factor [$(-1)^{\sma{F}}$] that a wavefunction accumulates upon a $2\pi$ rotation; $F{=}0$ (resp.\ ${=}1$) for integer-spin (resp.\ half-integer-spin) representations. The former case is useful in analyzing solids with negligible spin-orbit coupling, as we will exemplify with a case study of graphene in \s{sec:orbitsmutualT}.


The constraints in columns four and five are derived from the symmetry transformation of the Berry connection\cite{AA2014} and the one-form $\orb {\cdot} d\bk$ of \q{definerothoneform}. The latter may be expressed through Hamilton's equation [$\hbar \dot{\bk}{=}{-}|e| \bv {\times} \bB/\hbar c$] as ${-}\bM{\cdot}\bB dt/\hbar$, with  $\bM(\bk)$ an orbital moment that transforms under $g$ like the spatial components of a (3+1)-dimensional pseudovector
\e{ \bM\big|_{g\circ \bk} = (-1)^{s(g)}\det[\check{g}] \,\bar{g}\, K^{s(g)} \, (\check{g}\bM)\, K^{s(g)} \,\bar{g}^{-1}\big|_{\bk}. \la{gactsonMmultiband}}
$\check{g}$ here is a three-by-three orthogonal matrix representing the point-group component of $g$, and $K^s\bM K^s{=}\bM^*$ if $s{=}1$.

\subsection{Symmetry constraints on $\lambda$}\la{sec:constrainlambda}

From taking the determinant of each equation in column four, we derive constraints on $\lambda_a$ which are summarized in column six.  

 Three of six classes in [I,II-A] are  characterized by the reality condition  $e^{\sma{ i{\sum}_a\lambda_a}}{\in} \R$. This implies $\lambda{=}0$ or $\pi$ for a nondegenerate band, i.e., the orbit respectively encircles an even or odd number of Dirac points. $\lambda{=}\pi$ is exemplified by the Dirac surface state\footnote{While it is known that $\phi_{\sma{B}}{=}\pi$ for a time-reversal-invariant 2D Dirac fermion, it is not generally appreciated that $\phi_{\sma{R}}{=}\phi_{\sma{Z}}{=}0$.} of the $\Z_2$ topological insulator Bi$_2$Se$_3$;\cite{fu2007b,Inversion_Fu,moore2007,Rahul_3DTI}  an orbit on the Dirac cone is self-constrained by time-reversal ($T$) symmetry [class II-A,$u{=}0,s{=}1$]. $\lambda{=}\pi$ may result solely from crystalline symmetry, as exemplified by a mirror-symmetric orbit [class II-A,$u{=}1,s{=}0$] that encircles a surface Dirac cone of the topological crystalline insulator SnTe\cite{SnTe} [further elaborated in \s{sec:mirror}]. Finally, $\lambda{=}\pi$ can be protected by a composition of $T$ and crystalline symmetry [class I, $s{=}1$], as exemplified by Weyl metals having the same space group as WTe$_2$; the robustness of $\lambda$ depends sensitively on the field orientation with respect to the crystallographic symmetry axis, as elaborated in \s{sec:fieldorientation}.

For degenerate bands, the reality condition fixes ${\sum}_{\sma{a=1}}^{\sma{D}}\lambda_a$ to $0$ or $\pi$, but not $\lambda_a$ individually. This will be exemplified by our case studies of 3D Dirac metals and $\Z_2$ topological insulators, in \s{sec:spindegSOC} and \s{sec:bi2se3} respectively. In \s{sec:orbitselfT}, we further demonstrate  the reality constraint may be further strengthened for spin-degenerate orbits ($D{=}2$) that are self-constrained by $T$ symmetry -- into a zero-sum rule: $\lambda_1{+}\lambda_2{=}0$ mod $2\pi$. Moreover, if one considers the set of $\{\lambda\}$ contributed by all closed orbits in a $T$-symmetric solid, we propose that $\{\lambda\}$ comprise only pairs of $\lambda$, such that each pair individually satisfies a zero-sum rule, as elaborated in \s{sec:globalsumrule}. This global constraint on $\{\lambda\}$ might be viewed as an analog of the Nielsen-Ninomiya fermion-doubling theorem,\cite{NIELSEN1981} which states that there are always an equal number of left- and right-handed Weyl fermions on a lattice. 

The four II-B classes  are characterized by 
\e{\{\lambda^{\sma{i{+}1}}_{\sma{a}}\}_{\sma{a=0}}^{\sma{D-1}}{=}\{(\mo)^{\sma{u+s}} \lambda^{\sma{i}}_{\sma{a}}\}_{\sma{a=0}}^{\sma{D-1}}; \as i{\in} \Z_{\sma{L}}. \la{IIBconst}} 
Let us discuss the implications of the above equation for the three following cases (i-iii):\\

\noi{i} For $s{\neq}u$ and even $L$ (which necessarily holds if $u{=}1$), the minimal Landau-level degeneracy is $L/2$. As illustration, the two valley-centered Fermi surfaces ($L{=}2$) in the transition-metal dichalchogenide WSe$_2$ are mutually constrained by $T$ symmetry [class II-B,$u{=}0,s{=}1$]; the Landau levels are nondegenerate but exhibit a symmetric splitting, as elaborated in \s{sec:orbitsmutualT}.  \\

\noi{ii} For $s{\neq}u$ and odd $L$, the minimal Landau-level degeneracy is $L$;  if $D$ is also odd, it is necessary that $\{\lambda\}$ contains either $0$ or $\pi$. This is exemplified by three of four disconnected Fermi pockets ($L{=}3$) per valley of bilayer graphene (with trigonal warping);\cite{mccann_bilayergraphene,varlet_bilayergraphene} we refer to the three Fermi pockets that are  mutually constrained by the composition of $T$ and six-fold rotational symmetry $\rot_{6z}$ [class II-B, $u{=}0,s{=}1$]. For pedagogy, it is instructive to consider a model of bilayer graphene with spinless electrons, hence the orbit associated to each Fermi pocket is nondegenerate ($D{=}1$). The three-fold degenerate $\lambda{=}\pi$  reflects that each spinless pocket encircles a Dirac point.\cite{mccann_bilayergraphene,varlet_bilayergraphene}\\

\noi{iii} We would demonstrate that the same set of Fermi pockets, if equipped with multiple point-group symmetries, can belong to multiple symmetry classes in \tab{tab:tenfold}. For example, the above-mentioned three pockets are \emph{also} invariant under 
the three-fold rotational symmetry $\rot_{3z}$ ($u{=}s{=}0$). We may therefore apply \q{IIBconst} with $u{=}s$, which generally implies that the the minimal Landau-level degeneracy is $L$. Again, $L{=}3$ in the present example. We remind the reader that this three-fold degeneracy constraint was consistently implied by $T\rot_{6z}$; in addition, $T\rot_{6z}$ implies a stronger constraint that $\{\lambda\}$ contains either $0$ or $\pi$.





\section{Case studies}\la{sec:casestudies}

The utility of \tab{tab:tenfold} is illustrated in the following case studies of existing conventional and topological metals, which were introduced in the previous section [\s{sec:constrainlambda}].  

\subsection{Orbits mutually constrained by time-reversal symmetry: Application to graphene and transition metal dichalcogenides}\la{sec:orbitsmutualT}

Our first study encompasses materials with two time-reversal-related valleys in their band dispersion,\cite{xiaodong_tmd_review} as exemplified by graphene and monolayer WSe$_2$. We will demonstrate: (i) how orbits in time-reversal-invariant ($T$) solids can nevertheless develop a nonzero orbital magnetic moment, and (ii) the role of point-group symmetry in discretizing the Berry phase of valley-centered orbits.  

To explain (i), we point out that $T$ symmetry relates the magnetic moment of wavepackets at $\bk$ and $-\bk$ [cf.\ \q{gactsonMmultiband}]; this  mapping in $\bk$-space distinguishes the symmetry transformation of magnetic moments  in solids from that in atoms.  This allows for valley-centered orbits that are separated in $\bk$-space to \emph{individually} develop a magnetic moment -- since time reversal relates one valley to the other, the net magnetic moment must vanish. 

In more detail, let us consider a finite chemical potential (as measured from the Dirac point) where valley-centered orbits ($\frako_i$) are disconnected; we introduce here a valley index $i{\in}\{1,2\}$. The two orbits  are mutually constrained as $T{\circ}\frako_1{=}\frako_2$ [class II-B,$u{=}0,s{=}1$]; each of $\frako_i$ is also self-constrained as $T\inv{\circ}\frako_i{=}\frako_i$ [class I,s{=}1]. $T\inv$ imposes reality for the orbital component of wavefunctions at each $\bk$, leading to $\bM{=}0$ [cf.\ \q{gactsonMmultiband} with $D{=}s{=}1$ and $\check{g}{=}{-}I$]. In analyzing graphene (which has negligible spin-orbit coupling), it is useful to first neglect the spinor structure of its wavefunction and then account for the Zeeman effect after -- this was implicit in our previous assumption of $D{=}1$. Such `spinless' wavefunctions transform in an integer-spin representation ($F{=}0$ in \tab{tab:tenfold}); the corresponding $\cala$ in Tab.\ \ref{tab:tenfold} should be interpreted as \q{definenonabelianunitary} without the Zeeman term. 

The Berry phase of graphene is $\pi$, and therefore  $\tilde{\lambda}^i{:=}\phi^i_{\sma{R}}{+}\phi^i_{\sma{B}}{=}0{+}\pi$ for each valley labelled by $i{=}1,2$,  as is consistent with the reality constraint in [class I,$s{=}1$] of \tab{tab:tenfold}. We have added an accent to $\tilde{\lambda}^i$ to remind ourselves that it is the purely-orbital contribution to $\lambda^i$. Further accounting for the Zeeman contribution, 
\e{\lambda_{\pm}^i = \tilde{\lambda}^i \pm \pi \,\f{g_0}{2}\,\f{m_c}{m},\la{noSOC}} 
with $\tilde{\lambda}^i{=}\pi$, $\pm$ distinguishing two spin species, and  $m_c$ the cyclotron mass.  The symmetric splitting of $\lambda_{\pm}$ about $\pi$ implies an invariance under inversion: $\{\lambda_+^i,\lambda_-^i\}{=}\{{-}\lambda_+^i,{-}\lambda_-^i\}$ mod $2\pi$.  $T$ symmetry imposes the mutual constraint: 
\e{\{\lambda_+^1,\lambda_-^1\}=\{{-}\lambda_+^2,{-}\lambda_-^2\} \ins{mod} 2\pi,\la{Tmutual}} 
which follows from [class II-B,$u{=}0,s{=}1$] of \tab{tab:tenfold}. \qq{noSOC}{Tmutual} together imply $\{\lambda_+^1,\lambda_-^1\}{=}\{\lambda_+^2,\lambda_-^2\}$. To recapitulate, we have reproduced the well-known fact that graphene's Landau levels are valley-degenerate but spin-split by the Zeeman effect.

\begin{figure}[ht]
\centering
\includegraphics[width=7 cm]{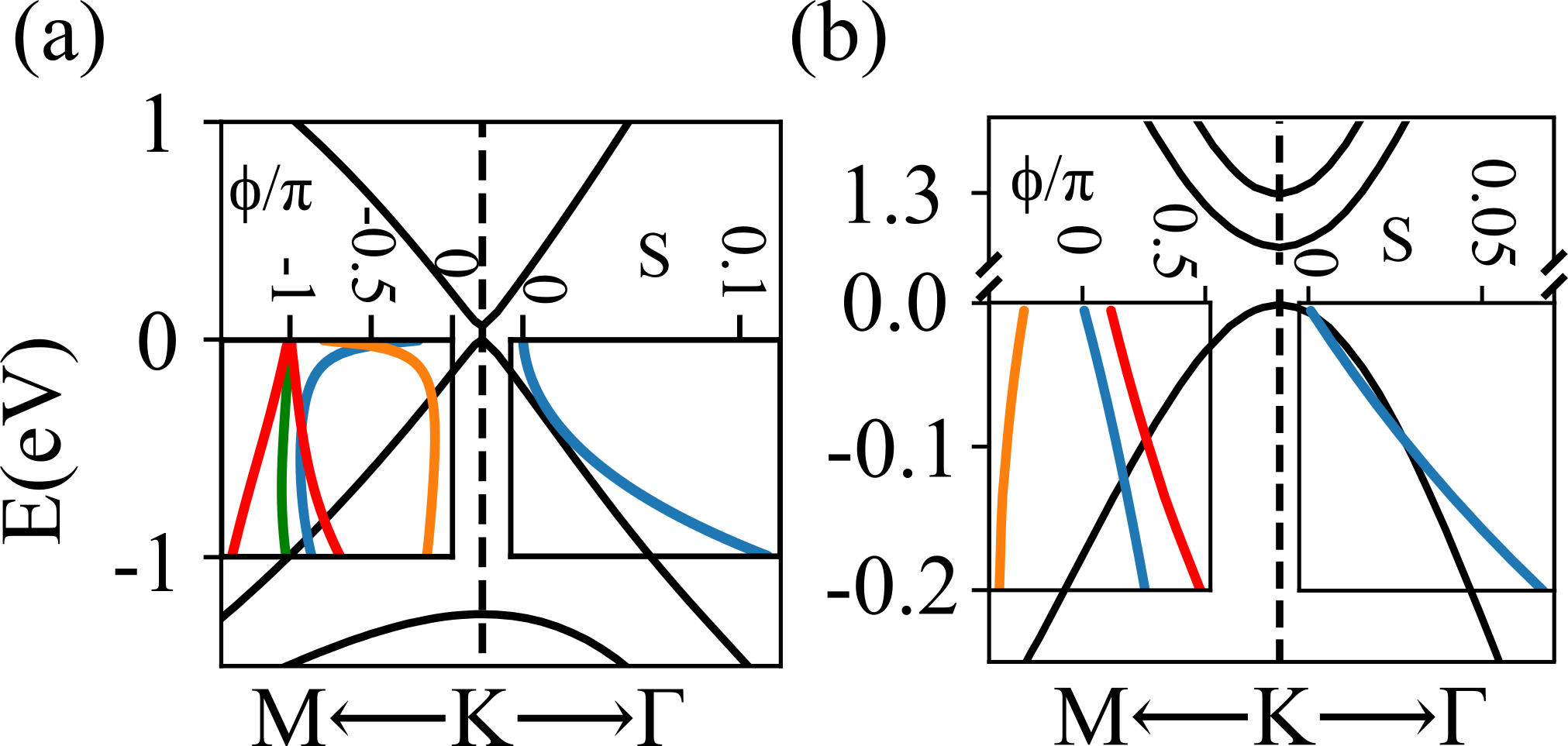}
\caption{(a) Characterization of a single valley of non-centrosymmetric graphene on hBN substrate, assuming a translation-invariant, commensurate\cite{Woods_incommensuratehBN} phase. Left inset: plot of $\phi_{\sma{B}}$ (blue), $\phi_{\sma{R}}$ (orange), $\tilde{\lambda}{:=}\phi_{\sma{B}}{+}\phi_{\sma{R}}$ (green) and $\lambda_{\pm}$ [cf.\ \q{noSOC}] (red) against energy. Right inset: plot of the orbit area ($S$) vs $E$.   (b) Characterization of a single valley of monolayer WSe$_2$. Left inset: plot of $\phi_{\sma{B}}$ (blue), $\phi_{\sma{R}}$ (orange), and $\lambda{:=}\phi_{\sma{B}}{+}\phi_{\sma{R}}{+}\phi_{\sma{Z}}$ (red) vs $E$. Right inset: plot of $S(E)$.  (a-b) are obtained by \emph{ab initio} calculations detailed in \app{app:abinitio}.}\label{fig:graphene_wse2}
\end{figure}

The valley degeneracy of the Landau levels may be split by breaking spatial-inversion ($\inv$) symmetry, e.g., with a hexagonal BN\cite{hBN_gap} or SiC\cite{Zhou_SiCsubstrate} substrate. In zero field, the $\inv$ asymmetry is predicted to produce a band gap of $53$ and $260$ meV respectively -- this may be interpreted as a nonzero Semenoff mass for the Dirac fermion.\cite{semenoff_mass} Since each valley-centered orbit is  no longer  self-constrained by $T\inv$, it develops a nonzero orbital moment (as was first noted in Ref.\ \onlinecite{DiXiao_Magneticmoment_valley}), as well as a non-integer $\phi^i_{\sma{B}}/\pi$. Consequently, \q{noSOC} remains valid with $\tilde{\lambda}^i$ deviating from $\pi$, as we confirm with a first-principles calculation in \fig{fig:graphene_wse2}(a).\footnote{The approximate equality of $\phi_{\sma{B}}{+}\phi_{\sma{R}}$ to $\pi$ may be understood from a linearized, two-band model  without next-nearest-neighbor hoppings\cite{fuchs_topologicalberryphase}} While $\{\lambda_+^i,\lambda_-^i\}$ (that is associated to one valley) is no longer invariant under inversion, we remark that the invariance persists for the full set:
$\{\lambda_+^1,\lambda_-^1,\lambda_+^2,\lambda_-^2\}$, owing to the mutual constraint by $T$ symmetry [cf.\ \q{Tmutual}].


The predicted Semenoff masses induced by substrates tend to be small --  it is instructive to compare graphene to WSe$_2$, a transition metal dichalcogenide with a large Semenoff mass due to the natural absence of inversion symmetry in its space group. WSe$_2$ is similar to graphene in that its low-energy bands (at zero-field) are also centered at two valleys, but differs from graphene in that its bands  are spin-split by spin-orbit coupling. \fig{fig:graphene_wse2}(b) illustrates our calculated $\lambda^1$ as a function of energy, with $i{=}1,2$ again a valley index, and $\lambda^1{=}{-}\lambda^2$ owing to $T$ symmetry; at $0.2$ eV below the band maximum,  $\lambda^1{-}\lambda^2{\approx}\pi/2$.






\subsection{Orbits self-constrained by mirror/glide symmetry: Application to topological crystalline insulators}\la{sec:mirror}


$\lambda/\pi{\in}\Z$ may originate from crystalline symmetry alone. This occurs for orbits that are self-constrained by a mirror/glide symmetry [IIA,$u{=}1,s{=}0$], which may arise in time-reversal-invariant and magnetic solids. Any circular orbit in [IIA,$1,0$] intersects a mirror/glide-invariant line at two points denoted by $\bk_a$ and $\bk_b$; at each point, the assumed-nondegenerate band transforms in either the even or odd representation of mirror/glide. The reality condition for $e^{i\lambda}$ in [IIA,$1,0$] of \tab{tab:tenfold} has a simple interpretation: $\lambda{=}0$ if the  representations at $\bk_a$ and $\bk_b$ are identical, and $\pi$ if the two representations are distinct.\cite{AA2014,LMAA,JHAA} The former is exemplified by a band that is nondegenerate at all $\bkp$ bounded by $\frako$, which implies that $\frako$ is contractible to a point -- due to continuity of the representation along the high-symmetry line, the representations at $\bk_a$ and $\bk_b$ must be identical.  $\lambda{=}\pi$ occurs iff there is an odd number of linear band touchings along the segment of the mirror line within $\frako$ -- at each touching (a Dirac point), the mirror/glide representation flips discontinuously. $\lambda{=}\pi$ is exemplified by the mirror-symmetric surface states of the SnTe-class\cite{SnTe} of topological crystalline insulators, as well as by glide-symmetric Dirac cones in band-inverted, nonsymmorphic metals.\cite{LMAA}



\subsection{Effect of field orientation on the crystalline symmetry of extremal orbits: Application to 3D Weyl fermion}\la{sec:fieldorientation}

\begin{figure}[ht]
\centering
\includegraphics[width=1\columnwidth]{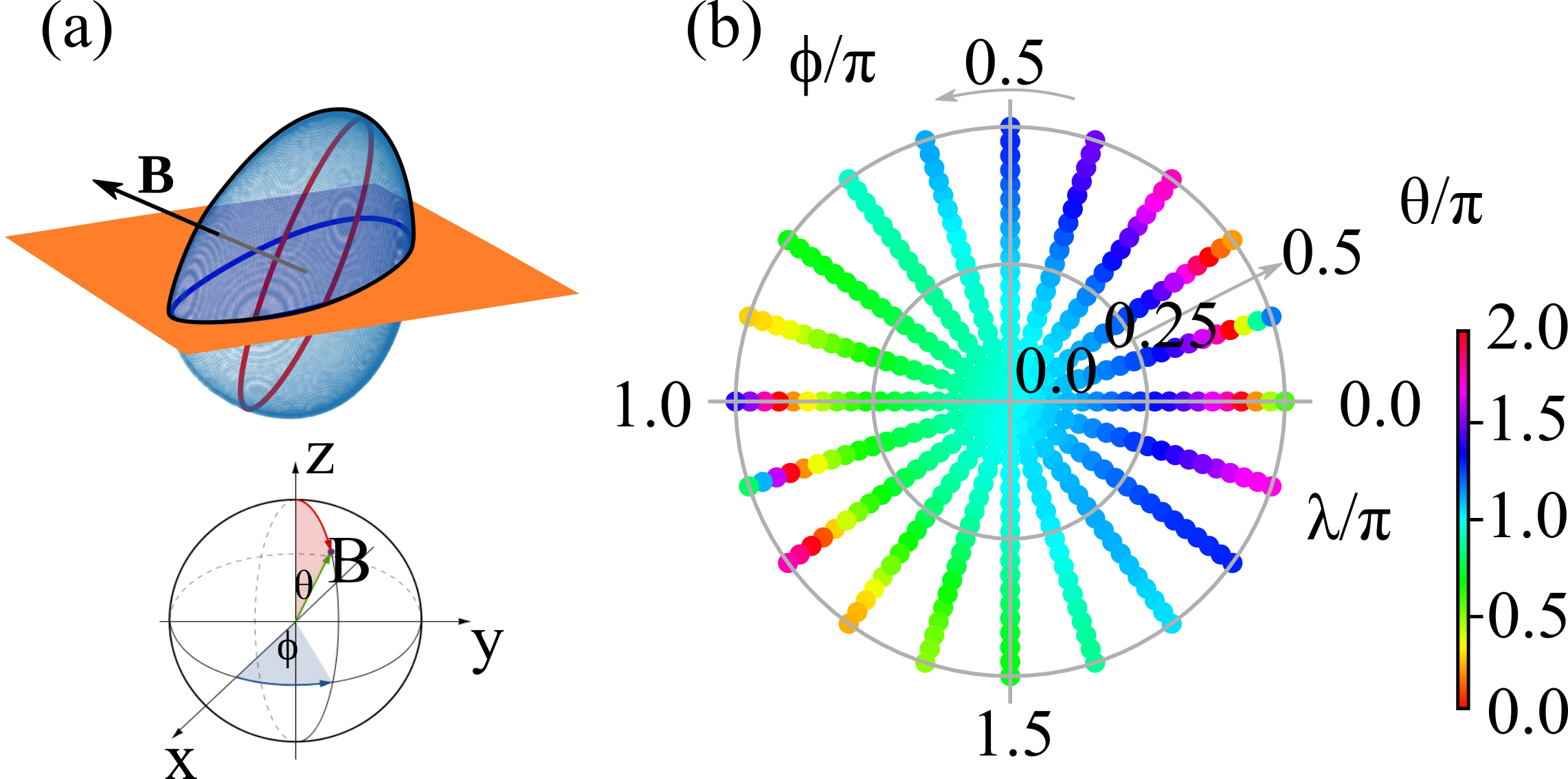}
\caption{\label{fig:WTe2-model}
(a) Fermi surface of a Weyl fermion centered on a generic wavevector in a high-symmetry plane; red circle indicates an extremal orbit that has no symmetry. (b) For a model of the Weyl fermion, we plot $\lambda$ as a function of the field orientation, which we parametrize by spherical coordinates illustrated in (a). 
}
\end{figure}

The previous examples demonstrate that  deforming the crystal structure offers a way to tune ${\lambda}$. For 3D solids, we may continuously tune between integer and non-integer $\lambda/\pi$ without explicitly breaking any symmetry -- by modifying the orientation of the field with respect to the crystal structure, we effectively modify the symmetry of the extremal orbit. For specific orientations, the extremal orbit may be invariant under a point-group symmetry that stabilizes integer-valued $\lambda/\pi$. We shall illustrate this with a 3D topological metal whose type-I Weyl points lie on a high-symmetry plane that is invariant under $T\scr_{2z}$ (the composition of time reversal with a two-fold screw rotation), as exemplified by strained WTe$_2$.\cite{WTe2Weyl} When the field is aligned parallel to the screw axis, the maximal orbit of a Weyl point lies within the $T\scr_{2z}$-invariant plane [class-I,$s{=}1$], and is characterized robustly by $\lambda{=}\pi$. As illustrated in Fig.\ \ref{fig:WTe2-model}, $\lambda$ deviates from $\pi$ when the field is tilted away, owing to the absence  of any symmetry for the tilted maximal orbit. 




\subsection{Orbits self-constrained by time-reversal symmetry}\la{sec:orbitselfT}

Let us consider orbits ($\frako$) which are self-constrained as $T{\circ}\frako{=}\frako$ [II-A,$u{=}0,s{=}1$]; these are orbits that lie in $T$-invariant cross-sections of the Brillouin torus, and encircle Kramers-degenerate points.  The contexts in which self-constrained orbits will be discussed  include:

\noi{a} $T$-invariant solids with negligible SOC,
 
\noi{b}  SO-coupled solids with both $T$ and spatial-inversion ($\inv$) symmetries, and 

\noi{c} SO-coupled solids with $T$ but without $\inv$ symmetry.\\

\noindent Cases (a) and (b) correspond to spin-degenerate bands. In the above cases, we would show respectively that:

\noi{a} The lack of SOC allows us to constrain the purely-orbital component ($\tilde{\lambda}$) of $\lambda_{\pm}$, where $\pm$ distinguishes the two spin species; recall that $\tilde{\lambda}$ and $\lambda_{\pm}$ differ only by the Zeeman splitting, as shown in \q{noSOC}. The orbit-averaged orbital moment vanishes,  and $\tilde{\lambda}{=}0$ reflects the trivial Berry phase of band extrema at $T$-invariant wavevectors. Combining $\tilde{\lambda}{=}0$ with \q{noSOC}, we obtain the following zero-sum rule: $\lambda_+{+}\lambda_-{=}0$ mod $2\pi$.

\noi{b} This zero-sum rule is also satisfied for spin-degenerate orbits in spin-orbit-coupled solids: $\lambda_1{+}\lambda_2{=}0$ mod $2\pi$. 

\noi{c} Both the orbital moment and Zeeman effect average out, and $\lambda{=}\pi$ reflects the nontrivial Berry phase associated to the Kramers-degenerate Dirac point at $T$-invariant wavevectors. \\

\noindent \emph{Demonstration} 

In all cases, the propagator $\cala[\frako]$ satisfies 
\e{\bar{T} \A^*\bar{T}^{\mo}=\A, \as \bar{T}\bar{T}^*=(-1)^{\sma{F}}\A^{\mo}, \as \bar{T}^{\mo} = \dg{\bar{T}}, \la{projectivealgebraAT}} 
with $\bar{T}K$ an antiunitary representation of $T$ [cf.\ row 4 of \tab{tab:tenfold} with $N{=}L{=}2$]. The second equation may be contrasted with the usual antiunitary representation (denoted $\tilde{T}K$) of $T$, which satisfies $\tilde{T}\tilde{T}^*{=}(-1)^{\sma{F}}$; the additional factor of $\A^{\mo}$ in \q{projectivealgebraAT} indicates that \q{projectivealgebraAT} represents an extension of the magnetic point group by contractible loop translations (represented by $\A$) in $\bk$-space,\cite{AALG_100} which generalizes a previous work on noncontractible loop translations.\cite{Cohomological} 

The first equation in \q{projectivealgebraAT} implies det$\,\A{=}e^{i{\sum}_{a=1}^{\sma{D}}\lambda_a}{=}{\pm}1$, with $D{=}2$ and $1$ for cases (b) and (c) respectively; in case (a), we should interpret det$\,\A{=}e^{i\tilde{\lambda}}{=}{\pm}1$. For cases (a) and (b), we may further restrict det$\,\A$ to ${+}1$ by the following argument: while preserving det$\,\A$, contract $\frako$ $T$-symmetrically  to an infinitesimal loop that encircles the $T$-invariant point. In the classes of (a-b), the band dispersion is extremized at $T$-invariant points, hence 
the band velocity $v(\bk)$ vanishes along the infinitesimal loop.\footnote{In principle, there may be a four-fold band touching (e.g., 3D Dirac point) at a $T$-invariant point that is accidental or imposed by crystalline symmetries. Note that such band touchings are not stable under a $T$-symmetric perturbation that breaks every other symmetry except, respectively, spin SU(2) symmetry in class (a), and $\inv$ symmetry in class (b). The value of det$\,\A$ cannot change due to this perturbation, since such value is quantized only by $T$ symmetry. The result of this argument is that we may always simplify the dispersion at the $T$-invariant point to a band extremum, and apply the proof in the main text. This perturbation argument is developed more carefully in \ocite{AALG_100}}  det$\,\cala$ is thus solely contributed by the non-geometric one forms which depend inversely on the velocity. Further applying the identity: det\,$\overline{\text{exp}}\int V(\bk)|d\bk|{=}$exp$\int$Tr$V(\bk)|d\bk|$ for any matrix $V(\bk)$, and the time-reversal constraints: Tr$\,\bM(\bk){=}{-}$Tr$\,\bM({-}\bk)$ [cf.\ \q{gactsonMmultiband} with $g{=}T$] and Tr$\,\bsigma(\bk){=}{-}$Tr$\,\bsigma({-}\bk)$, we derive the desired result. The above proof required path-ordering and matrix traces in case (b), where $\cala, \bM$ and $\bsigma$ are two-by-two matrices; such matrix operations are not needed in case (a).


 In case (c), we  contract $\frako$ $T$-symmetrically to an infinitesimal loop encircling the Kramers-degenerate Dirac point\footnote{By a $T$-symmetric perturbation, one may always simplify the dispersion to a Dirac point. This argument is analogous to that presented in the previous footnote.} -- since the velocity is finite in this limit, the non-geometric contribution to $\lambda$ vanishes; what remains is the $\pi$-Berry-phase contribution. This completes the demonstration. 

One implication of (a-b) for spin-degenerate orbits that are self-constrained by $T$ symmetry: the net phase offset $\Theta$ [cf.\  \q{defineTheta}], of two time-reversal-related fundamental harmonics, can only be $0$ or $\pi$. The former occurs if the two values of $\{\lambda\}$ are closer to $0$ than to $\pi$, and vice versa.


\subsection{Global constraint on $\{\lambda\}$ for time-reversal-symmetric solids}\la{sec:globalsumrule}

Let us impose a global constraint on the set [denoted $\{\lambda\}$] of all $\lambda_a^i$ that are contributed by closed orbits of bulk states in a $T$-symmetric solid. To clarify, for a $d$-dimensional solid, a `bulk state' is spatially extended in $d$ directions. By combining our results for orbits that are self-constrained [cf.\ \s{sec:orbitselfT}] and mutually constrained [cf.\ \s{sec:orbitsmutualT}] by time-reversal ($T$) symmetry, we would show that $\{\lambda\}$ comprise only of inversion-invariant pairs, i.e., pairs that are symmetric about zero. In our counting of `pairs', each closed orbit with energy degeneracy $D$ contributes $D$ values of $\lambda$, independent of whether any of these $D$ values are mutually degenerate, or degenerate with $\lambda$ from a distinct orbit. A corollary of this result is the global sum rule: $\sum \lambda {=}0$ mod $2\pi$. 


Let us first prove this claim for spin-degenerate orbits. All orbits in a $T$-symmetric solid are either self- or mutually constrained by $T$ symmetry. As proven in \s{sec:orbitselfT}, each self-constrained orbit contributes an inversion-invariant pair $\{\lambda,{-}\lambda\}$. Utilizing [class II-B,$u{=}0,s{=}1$] of \tab{tab:tenfold}, the net contribution of any $T$-related pair of spin-degenerate orbits is an inversion-invariant quartet $\{\lambda_1,\lambda_2,{-}\lambda_1,{-}\lambda_2\}$; this was exemplified by our case study of graphene in \s{sec:orbitsmutualT}. 

We would next prove the global constraint for spin-orbit-coupled solids lacking spatial-inversion symmetry [case (c) in \s{sec:orbitselfT}]. By similar reasoning as in the previous paragraph, each $T$-related pair of spin-\emph{non}degenerate orbits contributes $\{\lambda,{-}\lambda\}$. A self-constrained orbit necessarily contributes $\lambda{=}\pi$, as proven in \s{sec:orbitselfT}. Furthermore, self-constrained orbits always come in pairs; this follows because each self-constrained orbit encircles a $T$-invariant wavevector, and there are always $2^d$ such wavevectors in a $d$-dimensional lattice. This completes the proof. 

It is instructive to draw analogies between our result and the Nielsen-Ninomiya fermion-doubling theorem (as applied to $T$-invariant solids).\cite{NIELSEN1981} In full generality, the theorem states that there are always an equal number of left- and right-handed Weyl fermions on a lattice; a simple implication is that Weyl fermions come in pairs. If Weyl fermions exist at wavevectors which are \emph{not} $T$-invariant, they necessarily come in pairs due to $T$ symmetry mapping $\bk$ to $-\bk$. In the absence of point-group symmetry in the space group of the lattice, bands always come in pairs that  touch at isolated $T$-invariant wavevectors -- such Kramers-degenerate points are also Weyl points. That there are even number of Weyl fermions on a lattice thus complements our previous observation that there are an even number of Kramers-degenerate points. If the Fermi surface encircles Kramers-degenerate Weyl points instead of intersecting them, we recover our previous claim that self-constrained orbits come in pairs.   
    
To recapitulate, the global zero-sum rule for $\{\lambda\}$ is exemplified by all case studies of bulk orbits in this work; more generally it constrains the bulk oscillatory phenomena of all $T$-symmetric solids.

Let us extend our discussion to closed orbits contributed by surface states of  a 3D $T$-symmetric solid; these orbits lie in a 2D surface Brillouin zone. By a `surface state', we mean a state localized spatially to a surface that is translation-invariant in two directions; this surface lies at the interface between vacuum and a  bulk that is semi-infinite in the direction orthogonal to the surface.  If the solid is spin-orbit-coupled, surface bands are generically nondegenerate, and we would use nearly the same argument (given above) for spin-nondegenerate, bulk orbits; the sole difference is that the fermion-doubling theorem does not apply, and it is possible to have an odd number of self-constrained orbits (associated to an odd number of surface Dirac fermions). One implication is that $\chi_s{:}{=}\sum \lambda$ (summed over surface orbits) may equal $0$ or $\pi$. If the bulk of the solid is insulating, $\chi_s$ may be viewed as a $\Z_2$ index that classifies insulators from the perspective of its surface magnetotransport; this is equivalent to the $\Z_2$ strong classification\cite{fu2007b,Inversion_Fu,moore2007,Rahul_3DTI} of 3D insulators in Wigner-Dyson class AII. If the solid has negligible spin-orbit coupling, then the surface orbits satisfy the same type of global constraint as for bulk orbits -- to demonstrate this, one may utilize the above argument for spin-degenerate bulk orbits.


\subsection{Spin-degenerate orbits in spin-orbit-coupled, centrosymmetric metals: Application to the 3D Dirac metal Na$_3$Bi}\la{sec:spindegSOC}

A zero-sum rule also applies individually to each $T\inv$-invariant orbit ($\frako$ in  class [I,$s{=}1$]); this rule applies whether or not the orbit is self-constrained by $T$ symmetry. We have previously discussed such orbits in the context of graphene; here, we extend our discussion to $T\inv$-invariant, spin-degenerate orbits in SO-coupled solids. The reality of det$\,\A$ and contractibility of $\frako$ together imply $\lambda_1{+}\lambda_2{=}0$ mod $2\pi$.

\begin{figure}[ht]
\centering
\includegraphics[width=8cm]{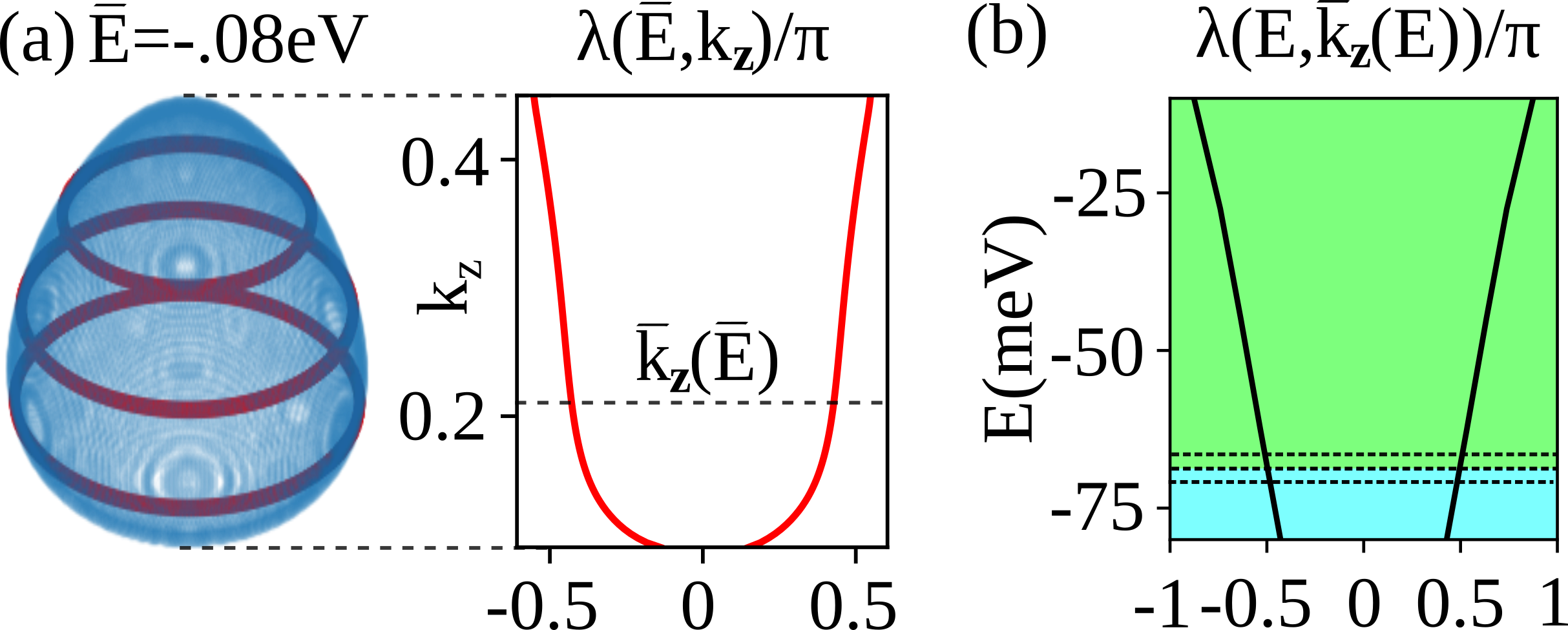}
\caption{\label{fig:Na3Bi}
Characterization of the 3D Dirac metal Na$_3$Bi. (a) Left: surface of fixed energy ($\bar{E}{=}{-}0.08$ eV measured from the 3D Dirac point); contours of fixed energy and $k_z$ are illustrated by red circles. Right: plot of $\{\lambda_a(\bar{E},k_z)\}_{a=1,2}$ for a field  parallel to the trigonal axis ($k_z$). $k_z$ is measured  in units where the reciprocal period ($0.64{\text{\AA}}^{\mo}$) is unity; the minimum (resp.\ maximum) $k_z$ on the plot corresponds to the south (resp.\ north) poles of the constant-energy surface. The extremal orbit occurs at $k_z=\bar{k}_z(\bar{E})$ indicated by a dashed line. (b) $\lambda_{1,2}(E,\bar{k}_z(E))$ vs $E$, for a range of $E$ below the Dirac point.}
\end{figure}

Let us apply this result to a 3D massless Dirac fermion -- a four-band touching point between conically-dispersing bands which are spin-degenerate at generic wavevectors (owing to $T\inv$ symmetry). We shall assume that the 3D Dirac point is \emph{not} centered at a $T$-invariant point, hence an orbit  at finite chemical potential (as measured from the Dirac point) is $T\inv$-invariant  but not $T$-invariant, e.g., such Dirac points in the topological metal Na$_3$Bi are stabilized by three-fold rotational symmetry.\cite{Na3BiZhijun} For a field aligned along the rotational (trigonal) axis, the equidistant splitting of $\lambda_a$  is illustrated in \fig{fig:Na3Bi}(a) for constant-$k_z$ orbits on a surface of constant energy ($0.08$eV below the Dirac point); $\lambda_{1,2}{\approx} {\pm}\pi/4$ for the maximal orbit (occurring at wavevector $\bar{k}_z$, as indicated by a dashed line on the plot). In \fig{fig:Na3Bi}(b), we further plot $\lambda_a(\mu,\bar{k}_z)$ for a range of $\mu$ below the Dirac point. The ab-initio calculation of $\lambda_a$ is detailed in \app{app:abinitio}.

Let us discuss the experimental implications for Na$_3$Bi. A single Dirac point contributes to the oscillatory magnetization a term of the form \q{oscmag3D}, with $D{=}2$ and $\lambda_{1,2}(\mu,\bar{k}_z)$ plotted in \fig{fig:Na3Bi}(b). Na$_3$Bi actually has two $T$-related Dirac points, which both lie on the rotation-invariant line through $\Gamma$.\cite{Na3BiZhijun} Assuming the Fermi surfaces of the two Dirac points are disconnected, the effect of the second Dirac point is simply to double the magnitude of the oscillations. This follows because the maximal orbits ($\frako_1,\frako_2$) on disconnected Fermi surfaces  are mutually constrained by $T$ symmetry, hence from class [II-B,$u{=}0,s{=}1$] of \tab{tab:tenfold}, $\{\lambda_1^1,\lambda_2^1\}{=}\{-\lambda_1^2,-\lambda_2^2\}$ mod $2\pi$. We remind the reader that $\lambda_1^i{=}{-}\lambda_2^i$ owing to $T\inv$ symmetry, hence in combination $\{\lambda_1^1,\lambda_2^1\}{=}\{\lambda_1^2,\lambda_2^2\}$.

\begin{figure}[ht]
\centering
\includegraphics[width=8.5cm]{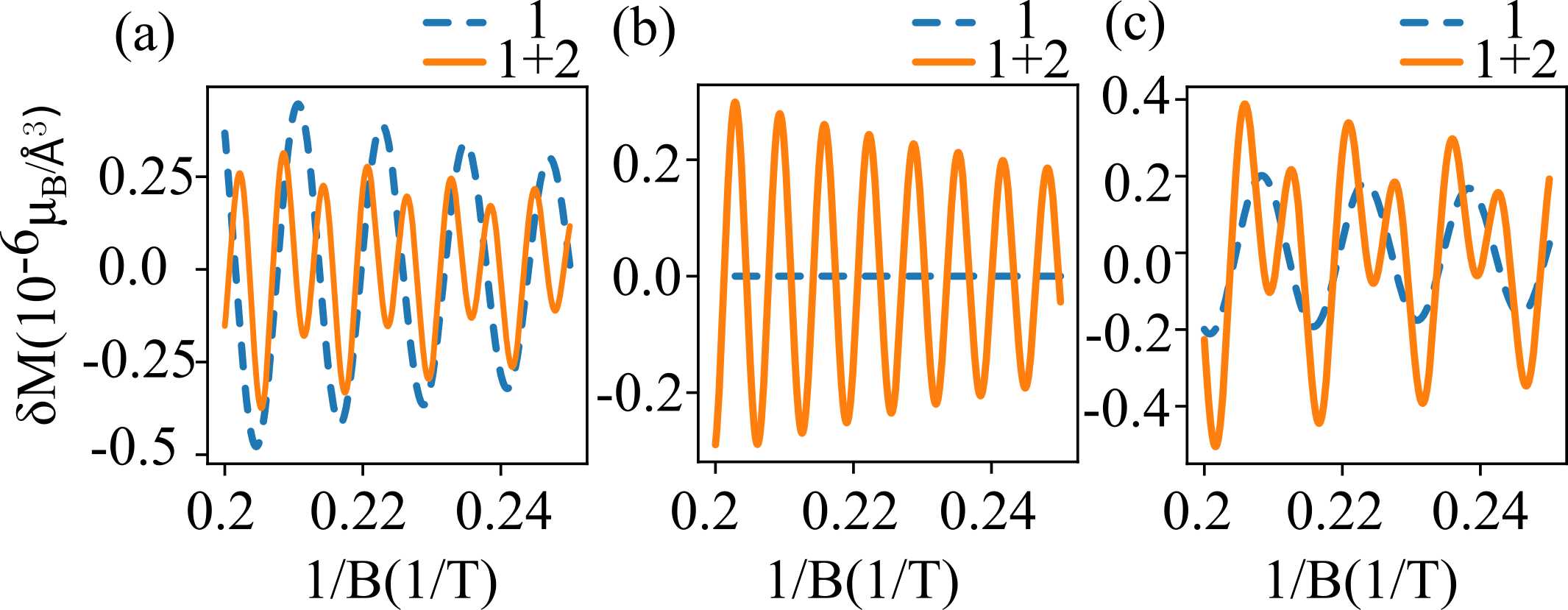}
\caption{\label{fig:Na3Bi-harmonics}
Magnetization oscillations calculated for Na$_3$Bi using \q{oscmag3D} at T=0.37K, $\tau{=}10^{-12}$s and Fermi energies (a) -72meV, (b) -69meV, and (c) -66meV. The corresponding $\lambda_{1,2}$ are indicated by dashed lines in \fig{fig:Na3Bi}(b). The sum of the two lowest harmonics [$r{=}1,2$ in \q{oscmag3D}], as illustrated by orange solid lines, faithfully represents the oscillations; however, accounting only for the fundamental harmonic ($r{=}1$, dashed blue line) is inadequate. The unit of magnetization is chosen as $10^{-6}\mu_{\sma{B}}/\ang^{\sma{3}}$, with $\mu_{\sma{B}}$ the Bohr magneton.}
\end{figure}

 To reiterate, the two Dirac points contribute identically to the oscillatory magnetization, which we plot in \fig{fig:Na3Bi-harmonics} for a temperature of 0.37 K and three representative Fermi energies. These plots are obtained from \q{oscmag3D}, which receive as inputs: (i) the ab-initio calculated band parameters $\{S, S_{zz},m_c,\lambda_{1,2}\}$ for the maximal orbit, as well as the Dingle lifetime $\tau{=}10^{-12}$s. For the range of $B$ plotted, $kT/\hbar\omega_c{\sim} 10^{-2}$ and $\omega_c\tau{\sim} 1$, which implies that the Dingle factor (rather than temperature smearing) is the limiting factor in observing higher harmonics; for the parameters considered, the third harmonic does not appreciably modify the plots. The second harmonic is especially transparent at $\mu_c{:}{=}{-}69$meV [\fig{fig:Na3Bi-harmonics}(b)], where $\lambda_1{=}\pi/2{=}{-}\lambda_2$ [cf.\ \fig{fig:Na3Bi}(b)] leads to the complete destructive interference of all time-reversed pairs of odd harmonics (including the fundamental harmonic) [cf.\ \q{oscmag3DD2}], i.e., the dHvA period is effectively halved. The point of destructive interference intermediates two energy intervals where the phase offset $\Theta$ [cf.\ \q{defineTheta}]  in the fundamental harmonic differs by $\pi$:  $\Theta{=}\pi$ for $\mu{>}\mu_c$ and closer to the Dirac point [cf.\ green shaded region in \fig{fig:Na3Bi}(b)]; $\Theta{=}0$ for $\mu{\leq} \mu_c$ [blue shaded region].  Our case study demonstrates that the experimental tunability\cite{satya_na3bi,Junxiong_Na3Bi,xiong2015signature} of $\mu$ in Na$_3$Bi should expose a wide range of $T$-symmetric interference patterns.




 For the parameter ranges in \fig{fig:Na3Bi} and \fig{fig:Na3Bi-harmonics}, the Fermi surfaces enclosing each Dirac point are indeed disconnected. As $\mu$ is increased beyond ${\sim} 10$ meV, the two Fermi surfaces intersect at a saddlepoint and merge into a dumbbell Fermi surface. For a field parallel to the trigonal axis, this merged Fermi surface has a single minimal orbit $\frako_3$ (encircling the saddlepoint), in addition to the aforementioned two maximal orbits. At slightly higher energy (${\sim} 20$ meV), the dumbbell transforms into an ellipsoid, and correspondingly the three extremal orbits merge into a single maximal orbit (denoted $\frako_4$). $\frako_3$ (and also $\frako_4$) is $T$-invariant [under class II-A,$u{=}0,s{=}1$] and $T\inv$-invariant [class I,$s{=}1$]; this exemplifies how a single orbit having multiple symmetries may fall under different classes in \tab{tab:tenfold}. Both classes impose (consistently) a zero-sum-rule constraint on $\{\lambda\}$.


\subsubsection{Comment on magnetotransport experiments of 3D Dirac metals}

We conclude this section by commenting on the interpretation of several magnetotransport experiments on 3D Dirac metals.\cite{he_transportCd3As2,xiang_sdHcd3as2,pariari_dhva_cd3as2,kumar_bi2se3,narayanan_cd3as2,zhaoyanfei_transportcd3as2} The experimental data in these works have only been fitted to the fundamental harmonic ($r{=}1$) in the traditional Lifshitz-Kosevich formula.\cite{shoenberg} The phase offset in this fitted fundamental harmonic is interpreted as $\phi_{\sma{B}}{-}\phi_{\sma{M}}{\pm} \pi/4$; $\phi_{\sma{B}}{=}\pi$ that is inferred from such a fitting is viewed as a smoking gun for a 3D Dirac metal. These interpretations have been  justified\cite{he_transportCd3As2,xiang_sdHcd3as2,pariari_dhva_cd3as2,kumar_bi2se3} by appealing to the \emph{one-band} Onsager-Lifshitz quantization rule [\q{rule3b} with $D{=}1$]. 


We propose an alternative interpretation that nevertheless makes sense of the robustness of their measurements, across different experimental groups and slightly-varying samples -- this robustness is solely a consequence of $T$ symmetry, rather than anything intrinsic to a 3D-Dirac Fermi surface. In other words, $\Theta{=}\pi$ should not be viewed as a smoking gun for a 3D Dirac metal.

From our perspective, spin-degenerate orbits in spin-orbit-coupled metals are semiclassically described by the multi-band quantization rule [cf.\ \q{rule3b}] and a generalized Lifshitz-Kosevich formula [cf.\ \qq{oscmag3D}{oscillatorydIdV}]. The phase corrections in these expressions encode the \emph{non-abelian} Berry, orbital-moment and Zeeman one-forms [cf.\ \q{definenonabelianunitary}]. Owing to the generic non-commutivity of the three one-forms in the path-ordered integral [cf.\ \q{definenonabelianunitary}], $\lambda_{\pm}$ cannot be decomposed into a sum of Berry, orbital-moment and  Zeeman phases.\footnote{ If $D{=}1$, the three one-forms commute, and therefore $\lambda$ may by decomposed. We remark that the $\pi$ Berry phase of a Weyl point originates from an integral of the \emph{abelian} Berry connection in a single-band subspace.} 

As described in \qq{oscmag3D}{defineTheta}, the phase offset ($\Theta$) of the summed fundamental harmonic, after substracting the Maslov and Lifshitz-Kosevich corrections, is a suitably-defined \emph{average} of $\lambda_{1,2}$. As proven in \s{sec:orbitselfT}, $\Theta$ is restricted to $0$ or $\pi$ for orbits that are self-constrained by $T$ symmetry. We believe that previous measurements\cite{he_transportCd3As2,xiang_sdHcd3as2,pariari_dhva_cd3as2,kumar_bi2se3,narayanan_cd3as2,zhaoyanfei_transportcd3as2}  of $\Theta$ have been misinterpreted as a measurement of the Berry phase. The convergence in measurements of several groups indicate the quantization of $\Theta$ due to $T$ symmetry. However, $\Theta{=}\pi$ is not an intrinsic property of 3D Dirac metals, as we have exemplified by our case study of  Na$_3$Bi. Moreover, we point out that $\Theta{=}\pi$ has been measured in conventional (i.e., non-topological) metals, even in the first metal (3D, single-crystal Bismuth) for which quantum oscillations were reported.\cite{SdH,dHvA} As far back as 1954, it was remarked by Dhillon and Shoenberg that: `For Bismuth good agreement was found by the theoretical formula except that the signs of the fundamental and odd harmonics had to be reversed.'\cite{dhva_Dhillon} Here, the `theoretical formula' refers to the traditional Lifshitz-Kosevich formula\cite{lifshitz_kosevich} without the $\Theta$ correction to the phase. In a later work, the quantization of $\Theta$ is implicit in the form of Shoenberg's spin `reduction factor',\cite{Dingle_I,shoenberg} which involves a \emph{phenomenological} g-factor in the presence of spin-orbit coupling. 

One contribution of this work is to present  an \emph{analytic} formula for the effective g-factor through the spectrum of \q{definenonabelianunitary}.\footnote{For small orbits encircling a high-symmetry wavevector, this effective g-factor may alternatively be calculated in the effective-mass approximation.\cite{gfactorBismuth_cohen_blount} For larger orbits, the effective-mass approximation is inappropriate, but \q{definenonabelianunitary} remains valid. Our method is essentially identical to previous works by Roth\cite{rothII}, Mikitik and Sharlai;\cite{Mikitik_quantizationrule} the only difference is that a basis choice (gauge) has been chosen in previous works such that their one-forms are traceless.\cite{AALG_100} In comparison, no gauge-fixing is assumed in \q{definenonabelianunitary}, and therefore \q{definenonabelianunitary} is the most natural object to calculate numerically.} Our proposed formula is calculable from \emph{ab initio} wavefunctions [as elaborated in \app{app:abinitio}], which allows for a quantitative comparison between theory and experiments. Another contribution of this work is the recognition that the quantization of $\Theta$ originates from $T$ symmetry; hence in magnetic materials, Shoenberg's spin `reduction factor' is not generally applicable, but our generalized Lifshitz-Kosevich formulae in \s{sec:extractlambda} remain valid.




We further emphasize that $\Theta$ is an incomplete characterization of the Fermi-surface orbit; a more complete characterization involves measuring the individual values of $\lambda_{1,2}$, which carry the complete non-abelian information about the Fermi-surface wavefunction. Two methods are available to measure $\lambda_{1,2}$: (i) where either of $\omega_c\tau {\ll} 1$ or $kT {\gg} \hbar \omega_c$ applies, one needs to measure both the amplitude and the phase offset of the fundamental harmonic; $\lambda_{1,2}$ is then extracted from the $r{=}1$ terms in \q{oscmag3D} and \qq{oscmag3DD2}{defineTheta}.\footnote{As mentioned in \s{sec:extractlambda}, this procedure requires an independent determination of $|S_{zz}|$.} (ii)  Where neither of the above inequalities apply, the experimental data should be fitted to the full harmonic expansion of \q{oscmag3D}, and $\lambda_{1,2}$ extracted accordingly. Note that the zero-sum rule for $\lambda_{1,2}$ implies that there is only one parameter to be fitted.


Going beyond general symmetry constraints, one may ask if $\Theta$ and $\lambda_{1,2}$ may be analytically calculated in a case-by-case basis. Given material-specific information about the magnetic point group of the 3D Dirac point, as well as the relevant symmetry representations that span the low-energy Hilbert space at the Dirac point, one may indeed determine $\Theta$ from a $\bk \cdot \bp$ analysis, as exemplified in the next section.

\subsection{Calculating $\lambda_{1,2}$ from $\bk \cdot \bp$ analysis: Application to  3D massive Dirac fermions in $\Z_2$ topological insulator Bi$_2$Se$_3$} \la{sec:bi2se3}

In certain symmetry classes, and for sufficiently small (but nonzero) chemical potentials ($\mu$) relative to the Dirac point, 3D massless Dirac fermions may exhibit  $\lambda_1{\approx}\lambda_2{\approx}\pi$; this would imply from \q{defineTheta} that $\Theta{=}\pi$. Precisely, we restrict $|\mu|$ to be small enough that a linearized $\bk\cdot \bp$ Hamiltonian is a good approximation -- how small $|\mu|$ needs to be to fulfil the above criterion depends on material-specific band parameters. As far as the linear approximation is valid, $\lambda_{1,2}$ that follows from the subsequent $\bk\cdot \bp$ analysis is approximately $\pi$, hence our use of $\approx$ throughout this section.  $\lambda_{1,2}$ should manifest in the dHvA oscillations for sufficiently weak field, such that  the semiclassical approximation  remains valid, i.e., $l^2S(\mu) {\gg}1$.\footnote{Note this implies a double constraint on $|\mu|$.}  


Let us demonstrate how $\lambda_{1,2}{\approx}\pi$ arises for 3D massless Dirac fermions centered at a wavevector ($\Gamma$) with a magnetic point group ($P$) that combines $T$ symmetry with the $D_{3d}$ point group.\cite{tinkhambook} $\lambda_{1,2}{\approx}\pi$  relies not just on $P$, but also on symmetries that are \emph{absent} in $P$; these additional symmetries emerge only at long wavelength  ($\bk {\rightarrow} \bze$), where the averaged-out crystal field has \emph{higher} symmetry than the magnetic point group. 

 Our study of $D_{3d}$ is motivated by the identical symmetry class of Bi$_2$Se$_3$, a well-known $\Z_2$ topological insulator.\cite{fu2007b,Inversion_Fu,moore2007,Rahul_3DTI} For the symmetry representations of Bi$_2$Se$_3$, the $\bk \cdot \bp$ Hamiltonian around $\Gamma$ (of the 3D Brillouin torus) assumes the form:\cite{zhang2009,chaoxingliu_modelhamiltonian}
\bal 
H_0 {=}  \hbar\big(v(k_x\sx{+}k_y \sy){+}wk_z\sz\big) {\otimes} \tx {+}M(\bk) I_{\sma{2\times 2}}{\otimes} \tz.\la{kdotpD3d}
\end{align}
to quadratic accuracy in $\bk$. Here, $\sigma_j$ and $\tau_j$ are Pauli matrices spanning spin and orbital (labelled $P1_z^+$ and $P2_z^-$)\cite{zhang2009} subspaces respectively, and $M(\bk){=}\Delta{+}b_1k_z^2{+}b_2(k_x^2{+}k_y^2)$ with $\Delta$, the Dirac mass, vanishing at the critical point between trivial and topological insulator. 

Assuming $\Delta{=}0$ for now, $[H_0(\bk),I_{\sma{2\times 2}}{\otimes}\tx]{=}O(\bk^2)$ is an emergent conservation law of the linearized $H_0$.  Block-diagonalizing $H_0$ with respect to $I_{\sma{2\times 2}}{\otimes}\tx{=}{\pm}1$, we derive two decoupled 3D Weyl Hamiltonians  $H_{\pm}$ with opposite chirality, each satisfying the time-reversal constraint $TH_{\pm}(\bk)T^{\mo}{=}H_{\pm}({-}\bk)$  with $T{=}i\sy K$ squaring to ${-}I$ in a half-integer-spin representation ($F{=}1$). Independent of the field orientation, the spin-degenerate extremal orbit effectively decouples to two nondegenerate $(D{=}1)$ extremal orbits which are individually $T$-invariant -- this implies from a previous demonstration [case (c) of \s{sec:orbitselfT}] that $\lambda_{1,2}{\approx}\pi$, which may be viewed as the Berry phase ($\phi_{\sma{B}}$) for each decoupled orbit. This would effectively imply  a single set of dHvA harmonics (indexed by $r$ in \q{oscmag3D}) with twice the usual amplitude, and a Lifshitz-Kosevich phase correction of ${-}\pi/4$ for a maximal orbit.  

Away from the critical point,  we prove in \app{sec:cancellation} that the degeneracy is Zeeman-split as $\lambda_{1,2}{\approx}\pi{\pm} \pi \Delta/mv^2$ with $m$ the free-electron mass,
for a field parallel to the trigonal axis ($k_z$). 
It is worth remarking that $\phi_{\sma{B}}$ reduces from $\pi$, and the phase ($\phi_{\sma{R}}$) associated to the orbital moment increases from $0$, however their sum ($\phi_{\sma{B}}{+}\phi_{\sma{R}}$) remains fixed to $\pi$.\footnote{ This robustness is an artifact of a restricted class of $\bk\cdot \bp$ models described in \app{sec:cancellation}.  The robustness is not generally valid at larger chemical potentials where the effect of other bands become significant. A similar observation was first found in \ocite{fuchs_topologicalberryphase} in the context of graphene with a Semenoff mass.}

In fact, naturally-occurring Bi$_2$Se$_3$ does not lie at the critical point, i.e., the corresponding 3D Dirac fermion is massive: we estimate $\Delta/mv^2{\approx}0.13$ utilizing fitted parameters from an ab-initio calculation in Ref.\ \onlinecite{zhang2009}; utilizing \q{defineTheta} with $|\Delta/mv^2|{<}0.5$, we further derive $\Theta{=}\pi$. This splitting of $\lambda$ manifests as two sets of harmonics in the dHvA oscillation, which should be measurable utilizing techniques outlined in \s{sec:spindegSOC}.  The tunability of $\Delta$ by doping\cite{Bi2se3_dope} or strain\cite{Bi2se3_strain,Young_Bi2se3_strain,Bi2se3_strain_expt} suggests that the dHvA oscillations are correspondingly tunable -- in particular, $|\Delta/mv^2|{=}0.5$ is the point of complete destructive interference for the fundamental harmonic; this point separates two regimes where $\Theta{=}0$ vs $\pi$.


There exist galvano-magnetic evidence that supports the quantization of $\Theta$. The fundamental Shubnikov-de Haas harmonic\cite{SdH,adams_holstein,magneticquantumeffects_Roth_argyres} in the transverse resistivity of Bi$_2$Se$_3$ has been fitted\cite{kazuma_bi2se3oscillations} as
\e{\Delta \rho_{xx} {\propto} \cos (l^2S{+}\Theta{+}\phi_{\sma{M}}{-}\pi/4) {\approx} \cos[l^2S{+}2\pi(0.4)],\la{fit}}
with the higher harmonics suppressed by temperature. The above proportionality may be derived from: (a) $\rho_{xx}{\approx}\sigma_{xx}/\sigma_{xy}^2$, which is valid for a large Hall angle $\sigma_{xy}{\gg}\sigma_{xx}$,  and (b) the proportionality between the oscillatory components of the transverse conductivity and the longitudinal magnetic susceptibility: $\Delta \sigma_{xx}\propto m_c \Delta \chi$;\cite{quantumcollision_kosevich_andreev,physicalkinetics_lifshitzpitaev} the cyclotron mass ($m_c$) is positive for the lone electron-like orbit in Bi$_2$Se$_3$,\cite{kazuma_bi2se3oscillations} and  the oscillatory susceptibility ($\Delta \chi$) is obtained from differentiating \q{oscmag3D} (with $D{=}2$) with respect to $B$ and keeping only the fastest oscillatory terms. The proportionality stated in (b) is valid for metals having only a single extremal orbit, and for sufficiently weak fields such that $\mu {\gg}\hbar \omega_c$.\footnote{In addition, there are auxiliary conditions such as $\hbar \omega_c{\ll}(kT \mu)^{\sma{1/2}}$ if $kT$ and $\hbar/\tau {\ll}\hbar \omega_c$.\cite{physicalkinetics_lifshitzpitaev} It has been assumed in \ocite{physicalkinetics_lifshitzpitaev} that impurity-induced scattering between spin-split Landau levels is absent; we are not certain that this assumption is generally valid in spin-orbit-coupled metals.} The fit in \q{fit} implies $\Theta{=}0.4{-}1/2{+}1/8{\approx}0$ to the accuracy of their fit.  It was further remarked in \ocite{kazuma_bi2se3oscillations}, without explanation, that the measured $\Theta{\approx}0$ was independent of the field orientation. To elaborate, four measurements, carried out at angles $0$, $\pi/4$, $\pi/3$ and $\pi/2$ relative to the trigonal axis in Bi$_2$Se$_3$, produced the same value for $\Theta$. Our explanation for this robustness: independent of the field orientation, the extremal orbit is invariant under time-reversal symmetry, which quantizes $\Theta$ to $0$ or $\pi$. In contrast, crystalline symmetries of an extremal orbit depend sensitively on the field orientation, as explained in \s{sec:fieldorientation}.

\section{Discussion}\la{sec:discussion}

In fermiology, the shape of the Fermi surface is deducible from the period of dHvA oscillations; here, we propose that the topology of the Fermi-surface wavefunctions are deducible from the phase offset ($\lambda_a$) of dHvA oscillations in 3D solids, as well as in fixed-bias oscillations of the differential conductance in scanning-tunneling microscopy. $\lambda_a$ manifests as the \emph{complete}, subleading [$O(1)$] corrections to the Bohr-Sommerfeld quantization rules for closed orbits without breakdown [cf.\ \q{rule3b}]; we have formulated $\lambda_a$ as eigen-phases of a propagator $\cala$ defined in \q{definenonabelianunitary}. 

In certain solids and for certain field orientations with respect to a crystal axis, ${\sum}_{\sma{a{=}1}}^{\sma{D}}\lambda_{\sma{a}}$ (with $D$ the degeneracy of the low-energy band subspace)  are topologically invariant under deformations of the zero-field Hamiltonian that respect the symmetry of the orbit, as well as preserves the global shape of the orbit. Precisely, globally-equivalent orbits correspond to a graph with a homotopy equivalence defined in Ref.\ \onlinecite{AALG_100}. To identify orbits with robustly integer-valued ${\sum}_{\sma{a{=}1}}^{\sma{D}}\lambda_{\sma{a}}/\pi$, as well as identify the degeneracy of Landau levels, we classify symmetric orbits into ten (and only ten) classes summarized in the first three columns of \tab{tab:tenfold}; the rest of the table describes the corresponding constraints on the propagator $\cala$. 



The results of this table remain valid if we substitute $\cala {\rightarrow}\W$ and $\lambda_a{\rightarrow}\phi_{\sma{B},a}$; here, $\W$ is defined as the unitary generated by the Berry connection, and may be viewed as the purely-geometric component of $\cala$; $\phi_{\sma{B},a}$ are defined as the eigen-phases of $\W$, and may be viewed as the non-abelian generalization\cite{wilczek1984} of the Berry phase. $\W$ provides a purely-geometric characterization of bands that is intimately related to the topology of wavefunctions over the Brillouin torus.\cite{zak1989,Cohomological,z2pack} While $\W$ (resp.\ $\phi_{\sma{B},a}$) generically differs from $\cala$ (resp.\ $\lambda_a$) due to the orbital moment and the Zeeman effect, $\W$ and $\A$ transform identically under symmetry, and therefore satisfy the same  constraints. In particular, if we define $\{\phi_{\sma{B}}\}$ as the set of $\phi_{\sma{B}}$  contributed by all bulk orbits in a $T$-symmetric solid, then $\{\phi_{\sma{B}}\}$ comprises only pairs of $\phi_{\sma{B}}$ which individually satisfy a zero-sum rule, in close analogy with the global constraint on $\{\lambda\}$ that is described in \s{sec:globalsumrule}.

Higher-order (in $|\bB|$) corrections to the Bohr-Sommerfeld quantization rule become relevant in higher-field experiments that intermediate the semiclassical and quantum limits;\cite{rothII,fischbeck_review} these corrections may be interpreted as zero-field magnetic response functions.\cite{Gao_Niu_zerofieldmagneticresponse} These corrections are accounted for in the generalized Lifshitz-Kosevich formulae [cf.\ \qq{oscmag3D}{oscillatorydIdV}] by the simple replacement $l^2S(E) {+} \lambda_a(E) {\rightarrow}l^2S(E) {+}\lambda_a(E,|\bB|),$ with $\lambda_a$ expandable asymptotically in powers of $|\bB|$; $\lambda_a(E,|\bB|{=}0){=}\lambda_a(E)$ is obtained from diagonalizing the propagator $\cala$ [cf.\ \q{definenonabelianunitary}]. It would be interesting to extend our symmetry analysis to the field-dependent component of the phase offset.


\begin{acknowledgments}
The authors thank Zhijun Wang and Ilya K. Drozdov for their expert opinions on 3D Dirac metals and tunneling spectroscopy. We also thank Yang Gao and Qian Niu for communicating their work on higher-order corrections in the quantization rule. We acknowledge support by  the Yale Postdoctoral Prize Fellowship (AA), NSF DMR Grant No.\ 1603243 (LG),  the Ministry of Science and Technology of China, Grant No.\ 2016YFA0301001 (CW and WD), and the National Natural Science Foundation of China, Grants No.\ 11674188 and 11334006 (CW and WD).\\
\end{acknowledgments}


\appendix

The appendix is organized as follows.\\

\noindent (\ref{app:dhva}) For bands of any degeneracy and symmetry, we derive the oscillatory magnetization and density of states in the de Haas-van Alphen and related effects.\\

\noindent (\ref{app:d3d}) We derive the eigen-phases ($\lambda_a$) of the propagator for the 3D massive Dirac fermion with $D_{3d}$ symmetry.\\
 
\noindent (\ref{sec:cancellation}) We prove the reality of the orbital component of $e^{i\lambda}$ for a class of two-band Hamiltonians, which generalizes a previous result for a certain class of massive Dirac fermions in Ref.\ \onlinecite{fuchs_topologicalberryphase}.\\

\noindent (\ref{app:abinitio}) We detail the \emph{ab initio} methods used in the calculation of \fig{fig:graphene_wse2}, and provide further details about the graphene-on-boron-nitride case study.\\

\noindent (\ref{app:tiltfield}) We provide the $\bk\cdot \bp$ model Hamiltonian (of a Weyl fermion) that was used in the calculation of \fig{fig:WTe2-model}.\\

\section{de Haas-van Alphen-type oscillations for bands of any degeneracy and symmetry}\la{app:dhva}

The following derivation of \qq{oscmag3D}{oscillatorydIdV} is a simple generalization of Roth's calculation\cite{rothII} for spin-degenerate bands in 3D solids. Consider a free-fermion system immersed in a magnetic field; we assume that near the Fermi energy, bands are $D$-fold degenerate, and the Landau levels $\{E_{a,j}\}_{a\in \{1,\ldots,D\}}$ are obtained semiclassically from \q{rule3b}. The grand-canonical potential is contributed by $D$ sets of sub-Landau levels:
\bal
 \Omega=\begin{cases} -kT  \cald_{\sma{3D}} \sum_{a=1}^D\int_{-\infty}^{\infty} dk_z \sum_{j=0}^{\infty} \phi(E_{a,j}), \\
-kT  \cald_{\sma{2D}} \sum_{a=1}^D\sum_{j=0}^{\infty} \phi(E_{a,j}),\end{cases}
\la{grandnonint}
\end{align}
for 3D and 2D metals respectively. The above expression involves
 \e{\phi(\var) := \log\left(1+e^{-\beta(\var-\mu)}\right), }
with $\beta{=}1/kT$, 
and the Landau-level degeneracy factor (per unit volume, or area): 
\e{  \cald_{\sma{3D}}:= \f{1}{4\pi^2 l^2}, \as \cald_{\sma{2D}}:=\f{1}{2\pi l^2}.\la{LLdeg}}
Utilizing the standard Poisson summation and following essentially Roth's calculation in \ocite{rothII}, the oscillatory component of $\Omega$ is derived as 
 \e{&\delta \Omega=   {(2\pi)^{1/2}}\f{kT \cald_{\sma{3D}}}{l|S_{zz}|^{1/2}} \lin
&\times \sum_{a=1}^{D}\sum_{r=1}^{\infty} \f{\cos\left[ r(l^2S+\lambda_a-\phi_{\sma{M}}) \pm \pi/4\right]}{r^{3/2} \,\text{sinh}\,(2\pi^2kT/\hbar \omega_c)}\bigg|_{\mu,\bar{k}_z},\la{oscillatorygrand}}
for 3D metals, and 
\e{\delta \Omega\eq  2kT \cald_{\sma{2D}} \sum_{a=1}^{D} \sum_{r=1}^{\infty} \f{\cos{\left[r(l^2S+\lambda_a-\phi_{\sma{M}})\right]}}{r \,\text{sinh}\,(2\pi^2kT/\hbar \omega_c)}\bigg|_{\mu},}
for 2D metals. Employing the thermodynamic definition of magnetization as $M{=}{-}\partial \Omega/\partial B$, and keeping only the fastest oscillatory term in the semiclassical limit ($l^2S {\gg}1$), the oscillatory component of the magnetization is derived as \qq{oscmag3D}{oscmag2D} without the Dingle factor.\cite{Dingle_collisions} 

Employing the following relation:\cite{shoenberg} $g(\mu){=}{-} \partial^2\Omega/\partial \mu^2|_{T=0}$ between the density of states and the zero-temperature grand-canonical potential, we differentiate  \q{oscillatorygrand} [at zero temperature, and inclusive of the Dingle factor] twice to obtain the oscillatory component of the 3D density of states:
 \e{&\delta g(E)=   \f1{\sqrt{2}\pi^{3/2}} \f1{\hbar\omega_c} \f1{l^3|S_{zz}|^{1/2}} \lin
&\times \sum_{a=1}^{D}\sum_{r=1}^{\infty}e^{-\tf{r\pi }{\omega_c\tau}}\f{\cos\left[ r(l^2S{+}\lambda_a{-}\phi_{\sma{M}}) {\pm} \pi/4\right]}{r^{1/2} }\bigg|_{E,\bar{k}_z}.\la{oscillatorydos}}
By convolving this quantity with the derivative of the Fermi-Dirac distribution (as in \q{defineTbroadenedDOS}), we derive that the oscillatory component of $\calg$ is \q{oscillatorydIdV}. This derivation is aided by the identity\footnote{A derivation can be found in the subsection titled `The Sommerfeld expansion' in \ocite{lecturenotes_arovas}}
\e{-\int_{-\infty}^{\infty}d\var \,f_T'(\var-\mu-E)\,\Phi(\var)=\f{\pi D}{\sin \pi D}\Phi(\mu+E),}
with $D{:}{=}kT(\partial/\partial\mu)$; the above differential operator is defined through the Taylor expansion: $x/\sin x{=}1{+} x^2/6{+}7x^4/360{+}\ldots$, which is well-known in its application to the Sommerfeld expansion. Applying the above identity with $\Phi{=}\delta g$, and retaining only the quickest oscillatory terms  [i.e., only the terms derived from $D$ acting on the cosine function:
\e{ \f{\pi D}{\sin \pi D}\cos(\Gamma)=\f{\pi kT d\Gamma/d\mu}{\text{sinh}(\pi kT d\Gamma/d\mu)}\cos(\Gamma)\;\bigg],} 
we arrive at the desired result.

\section{3D massive Dirac fermion with $D_{3d}$ symmetry}\la{app:d3d}

The linearized $\bk\cdot\bp$ Hamiltonian from \q{kdotpD3d} 
\bal 
H(\bk) \eq  \hbar\big(v(k_x\sx{+}k_y \sy){+}wk_z\sz\big) {\otimes} \tx +\Delta I_{\sma{2\times 2}}{\otimes} \tz\la{kdotpD3dsupp}
\end{align}
with a nonzero mass $\Delta$. This Hamiltonian has an emergent reflection symmetry:
\e{  \breve{\mir}_zH_0(\bkp,k_z)\,\breve{\mir}^{\mo}_z=H_0(\bkp,-k_z), \as \breve{\mir}_z=-i\sz \tz, \la{emergentmz}} 
which is broken by terms which are cubic in $\bk$.\cite{chaoxingliu_modelhamiltonian} If the field is oriented parallel to the trigonal axis $(k_z)$,  the maximal orbit lies at $k_z=0$ owing to the just-mentioned symmetry. It is here that $[H(\bkp,0),\breve{\mir}_z]=0$, and  we may block-diagonalize $H$  with respect to even and odd representations of $\breve{\mir}_z$:
\e{ H_{\pm}=\hbar v(k_x \gamma_1+k_y\gamma_2)\pm \Delta \gamma_3.}
$H_{\pm}$ describe two 2D massive Dirac fermions with opposite chirality. It is known for each of $H_{\pm}$ that $\phi_{\sma{R}}{+}\phi_{\sma{B}}{=}\pi$ is independent of the Semenoff mass ($\Delta$);\cite{fuchs_topologicalberryphase} we rederive this result in our language in  in \s{sec:cancellation}, where we further clarify the applicability of their result to more general two-band Hamiltonians.  What remains is to calculate  the Zeeman contribution to $\lambda$. Employing that the spin operator $S_z$ is represented $\hbar \gamma_z/2$, and 
\e{v^{\perp} = \f{v^2\hbar |\bk|}{\sqrt{(\hbar v \bk)^2+\Delta^2}},}  the Zeeman phase is
\e{ \phi_{\sma{Z}}=  \f{g_0\hbar}{4m}\int_{\frako} \f{|d\bk|}{v^{\perp}} \f{\Delta}{\sqrt{\Delta^2+(\hbar v\bk)^2}} 
 \approx \pi \f{\Delta}{mv^2}.}
Employing the following parameters from Ref.\ \onlinecite{zhang2009}: $\Delta=0.28$eV, $v=6.2\times 10^5 ms^{-1}$, we obtain $\phi_{\sma{Z}}\approx 0.13\,\pi$.

\section{Reality of the orbital component of $e^{i\lambda}$ \label{sec:cancellation}}

For a general dispersion $\epsilon(\mathbf{k})$, the area of constant
energy contour is a function of $\epsilon$: $S(\epsilon)$. For a
general scalar valued function $f(\mathbf{k})$, we define a $F(\epsilon):=\int_{S(\epsilon)}f(\mathbf{k})d^{2}\mathbf{k}$.
Then the average of $f$ on the contour is defined by:
\begin{align}
\overline{f} & :=\frac{dF}{d\epsilon}/\frac{dS}{d\epsilon}=\int_{\leftturn}\frac{f}{v_{x}}dk_{y}/\frac{dS}{d\epsilon}.
\end{align}
Thus 
\begin{align}
\phi_{R} & =\int_{\rightturn}-\frac{l^{2}BM}{v_{x}}dk_{y} =l^{2}B\overline{M}\frac{dS}{d\epsilon},
\end{align}
where $M$ is orbital magnetization. The sign change comes from a change in orientation in path. For two band Hamiltonian in the form of
\begin{equation}\label{eq:two-band-hamiltonian}
H(\mathbf{k})=\left(\begin{array}{cc}
\Delta & r(\mathbf{k})e^{i\theta(\mathbf{k})}\\
r(\mathbf{k})e^{-i\theta(\mathbf{k})} & -\Delta
\end{array}\right),
\end{equation}
energy dispersion are symmetric with respect to zero. Here, $r$ is real and $\Delta$ is $\mathbf{k}$ independent. In this case, orbital magnetization are simply related to Berry curvature by $M=\frac{e}{\hbar c}\Omega\epsilon$, where Berry curvature is defined by $\mathbf{\Omega} = \nabla\times\bmx(\bk)$. Then
\begin{align}
\phi_{R} & =l^{2}\frac{eB\epsilon}{\hbar c}\overline{\Omega}\frac{dS}{d\epsilon} =\overline{\Omega}\epsilon\frac{dS}{d\epsilon} =\epsilon\frac{d\phi_B}{d\epsilon},
\end{align}
 and 
\begin{align}
\phi_{R}+\phi_{B} & =\epsilon\frac{d\phi_B}{d\epsilon}+\phi_B =\frac{d\epsilon\phi_B}{d\epsilon}.
\end{align}
$\frac{d\epsilon\phi_B}{d\epsilon}$ is evaluated to be $\pm\pi\int_{\rightturn}d\mathbf{k}\cdot\nabla_{k}\theta$ for conduction band and valence band respectively, which is fixed to $0$ or $\pi$ (mod $2\pi$).\\

Any two-band Hamiltonian that can be transformed into Eq. (\ref{eq:two-band-hamiltonian}) (i.e., with at least one $\bk$-independent multiplicative coefficient of a Pauli matrix) is also characterized by real $e^{i(\phi_R+\phi_B)}$. We propose a sufficient condition for the existence of this transformation, which we denote by $U$ in what follows.\\

\noindent \emph{Sufficient condition} In general, for any three-by-three special orthogonal matrix ($R$), there exists a $\bk$-independent unitary $U$ such that
\begin{equation}
U^\dagger\mathbf{d}\cdot\bsigma U = (R(U)\mathbf{d})\cdot \bsigma,\as R(U)\in\text{SO}(3),
\end{equation}
where $\bd(\bk)$ is a real three-vector, and $\bsigma=(\sx,\sy,\sz)$ are Pauli matrices.\cite{tinkhambook} Let us separate each component of $\bd$ into its $\bk$-independent and -dependent parts:  $d_j(\bk)=d_j^0 + d_j^1(\bk)$. If $d^1_x, d^1_y, d^1_z$ are linearly-dependent functions of $\mathbf{k}$, such a transformation is always possible. Indeed, suppose $d^1_y$ and $d^1_z$ are linearly independent, and by assumption we might express $d^1_x=\alpha d^1_y+\beta d^1_z$ for $\alpha,\beta \in \R$. Then the desired basis transformation maps 
\e{ d_3 \rightarrow d_3'=\tf{1}{1+\alpha^2+\beta^2} (-d_1+\alpha d_2 +\beta d_3)}
such that $d_3'$ is $\bk$-independent. The above coefficients of $d_j$ may be viewed as a real three-vector with unit norm. Therefore, the transformation from $d_3\rightarrow d_3'$ might be viewed as a special-orthogonal rotation ($R$) of a Euclidean coordinate system. Since $R$ exists that is $\bk$-independent, so thus $U$. This completes the demonstration.   \\

\noindent As an example, if $\bd(\bk)$ is at most linear in $\bk$, with $\bk=(k_x,k_y)$ restricted to a Brillouin two-torus, the above condition is satisfied owing to the existence of only two linearly-independent terms: $k_x$ and $k_y$, and therefore $\phi_{R}+\phi_{B}$ is fixed to either $0$ or $\pi$. This class of two-band, linearized Hamiltonians includes, as a special case, the massive-Dirac Hamiltonian [cf.\ \q{eq:two-band-hamiltonian}] first discussed in Ref.\ \onlinecite{fuchs_topologicalberryphase}.\\

\section{\emph{ab initio} calculations leading to Fig. 1 and further details about the graphene-hBN case
study} \la{app:abinitio}

Our \emph{ab initio} calculations are carried out within the framework of density functional theory, as implemented in quantum espresso software package\cite{giannozzi_quantum_2009}. Norm-conserving potentials\cite{schlipf_optimization_2015} and Perdew-Burke-Ernzerhof (PBE)  exchange-correlation functional\cite{perdew_generalized_1996} are used to describe electron-ion, electron-electron interactions respectively. Most of the physical quantities are obtained by Wannier interpolation, where maximally-localized Wannier functions are obtained by wannier90 code\cite{mostofi_updated_2014}. For the graphene-boron-nitride heterostructure, vdW correction\cite{reckien_implementation_2012} is further included. Calculation of the propagator $\mathcal{A}$ has been coded into python package owl,\cite{owlpaper,owllink} which has been developed for general-purpose Wannier interpolation.

When graphene is placed on a boron-nitride substrate, the Dirac point splits with an energy gap of approximately
60 meV. The cyclotron mass $m_c(E)$ [defined as $(\hbar^2/2\pi)dS/dE$, with $S(E)$ the k-space area of the constant-energy band contour], when evaluated at the band edge, is approximately
1000 eV$/c^2$ with $c$ the speed of light; this mass is approximately 0.002 times the free-electron mass.

\section{$k\cdot p$ model of Weyl fermions with $C_{2T}$ symmetry} \la{app:tiltfield}

In spinor basis, $T$ is represented by $i\sigma_yK$, $C_{2z}$ (two-fold rotation about $\vec{z}$) by $i\sigma_z$; we have picked a basis where the orbital component of the basis functions transforms trivially under $C_{2z}$. We consider a $\bk \cdot \bp$ Hamiltonian expanded around a generic point on the $k_z=0$ plane; it has the symmetry
\e{ C_{2T}H(\bk)C_{2T}^{-1}=H(k_x,k_y,-k_z)}
with $C_{2T}:=TC_{2z}$ represented by $i\sigma_x K$. A symmetry-allowed Hamiltonian assumes the form
\e{
H \eq (Ak_x+Bk_y)I+(ak_x+ck_y+e(k_x^2+k_y^2))\sigma_x \lin
&+\; bk_x\sigma_y+dk_z\sigma_z.}

We do not exhaust all possible quadratic terms -- one is sufficient to demonstrate the anisotropy of $\lambda$. One may verify that the quadratic term causes $\phi_R+\phi_B$ to deviate from $\pi$ (see \app{sec:cancellation}). Furthermore, $k_y\sigma_y$ is not present in the Hamiltonian since it can be eliminated by a rotation in the $k_x-k_y$ plane. We choose the parameters to be A = 0.2, B = 0.2, a = 1, c = 0.3, b = 1, d = 0.2, e = 0.5 (eV$\cdot$\AA) to produce Fig. \ref{fig:WTe2-model} in the main text. In Fig. \ref{fig:WTe2-model}, constant energy contour is evaluated at -0.02eV.

\bibliography{bib_June2017}

\end{document}